\documentclass[fleqn,usenatbib]{mnras}
\usepackage{newtxtext,newtxmath}

\usepackage[T1]{fontenc}
\usepackage{ae,aecompl}

\usepackage{graphicx}	
\usepackage{amsmath}	

\usepackage{tikz}
\usepackage{listings}
\usepackage{caption}
\usepackage{animate}
\usepackage{subcaption}
\usetikzlibrary{positioning}
\usepackage[utf8]{inputenc}
\usepackage{setspace}
\usepackage{bm}
\usepackage{caption}
\usepackage{float}
\usepackage{tabularx}
\usepackage{verbatim}
\usepackage{placeins}
\usepackage{color}
\usepackage{ulem}

\newcommand\rinf{r_{\rm inf}}

\newcommand\ah{a_{\rm h}}

\newcommand{\epsy}{\epsilon_y}
\newcommand{\epsz}{\epsilon_z}

\title[Binary flip]{Formation of counter-rotating and highly eccentric massive black hole binaries in galaxy mergers}

\author[I. T. Nasim et al.]{%
Imran Nasim$^{1}$,\thanks{E-mail: i.nasim@surrey.ac.uk (KTS)}
Cristobal Petrovich$^{2,3}$,
Adam Nasim$^{4}$,
Fani Dosopoulou$^{5,6}$, \newauthor
Fabio Antonini$^{1,7}$
\\
$^{1}$ Department of Physics, University of Surrey, Guildford, GU2 7XH, Surrey, UK\\
$^{2}$ Steward Observatory, University of Arizona, 933 N. Cherry Ave., Tucson, AZ 85721, USA\\
$^{3}$ Instituto de Astrofísica, Facultad de Física, Pontificia Universidad Católica de Chile, Casilla 306, Santiago 22, Chile\\
$^{4}$ Department of Mathematics, University of Surrey, Guildford, GU2 7XH, Surrey, UK\\
$^{5}$Princeton Center for Theoretical Science, Princeton University, Princeton, NJ 08544, USA\\
$^{6}$Department of Astrophysical Sciences, Princeton University, Princeton, NJ 08544, USA\\
$^{7}$ Gravity Exploration Institute, School of Physics and Astronomy, Cardiff University, Cardiff, CF24 3AA, United Kingdom\\
}

\date{}
\pubyear{2020}

\begin{document}
\label{firstpage}
\pagerange{\pageref{firstpage}--\pageref{lastpage}}
\maketitle

\begin{abstract}
Supermassive black hole (SMBH) binaries represent the main target for missions such as the Laser Interferometer Space Antenna and Pulsar Timing Arrays. The understanding of their dynamical evolution prior to coalescence is therefore crucial to  improving detection strategies and for the astrophysical interpretation of the gravitational wave data. In this paper, we use high-resolution  $N$-body simulations to model the merger of two equal-mass galaxies hosting a central SMBH.  In our models, all binaries are initially prograde with respect to the galaxy sense of rotation. But, binaries that form with a high eccentricity, $e\gtrsim 0.7$, quickly reverse their sense of rotation and become almost perfectly retrograde at the moment of binary formation. The evolution of these binaries proceeds towards larger eccentricities, as expected for a binary hardening in a counter-rotating stellar distribution. Binaries that form with lower eccentricities remain prograde and  at comparatively low eccentricities. We study the origin of the orbital flip by using an analytical model that describes the early stages of binary evolution. This model indicates that the orbital plane flip is due to the torque from the triaxial background mass distribution that naturally arises from the galactic merger process. Our results imply the existence of a population of SMBH binaries with  a high eccentricity
and could have significant implications for the detection of the gravitational wave signal emitted by these systems.
\end{abstract}




\section{Introduction}
\label{sec:intro}
In the standard $\Lambda$CDM cosmological model, supermassive black hole (SMBH) binaries are  expected to form as a natural product  of galaxy mergers \citep[e.g.,][]{begelman1980}. 
The gravitational wave radiation emitted by such binaries during their inspiral and coalescence is the main target for the future Laser Interfermoter Space Antenna  \citep[LISA;][]{LISA_AMARO_SEOANE2017}  and for  Pulsar Timing Array (PTA) searches \citep{nanogravpta2015,reardonpta2016}.

 Strategies for gravitational wave searches focus mostly on quasi-circular motion, with the motivation that the orbit is expected to circularise well before the binary enters the frequency band of the gravitational wave detector (i.e., $\gtrsim 10^{-4}$Hz for LISA). When the previous dynamical evolution of the binary is considered, however,  the residual eccentricity inherited from the previous environment-dominated phase  can be substantial, and  can significantly affect both the gravitational wave signal and the merger time scale of the binary \citep[e.g.,][]{preto2011,GM2012,Nasim2020}. 
Large  binary eccentricities can also affect significantly the power spectrum of the gravitational wave background radiation from SMBH binaries \citep{2007PThPh.117..241E}.
 The understanding of the formation and dynamical evolution of SMBH binaries in merging galaxies is therefore a key to  improving data analysis and detection strategies, for the design of future gravitational wave observatories, and to constrain the physics of SMBH evolution from gravitational wave data.

The evolution of a SMBH binary towards coalescence can be divided into three separate phases \citep[e.g.,][]{begelman1980,merritbook2013}. The first phase of evolution is via the process of dynamical friction \citep[e.g.,][]{2011MNRAS.411..653J,2012ApJ...745...83A,DA2017}. During this first phase, the SMBHs sink towards the centre of the stellar system that resulted from the merger of the two progenitor galaxies, leading to the formation of a gravitationally bound pair. Further orbital decay through dynamical friction occurs until the binary becomes ``hard'' (see equation~(\ref{eq: hard_binary_sep}) below). At this point, dynamical friction becomes  less efficient and  subsequent evolution is driven by strong dynamical interactions of the binary with surrounding  stars
\citep{hills1983,quinlan1996,sesana2006}.
When the SMBH binary  orbit reaches a separation  of roughly one milli-parsec,  energy loss by gravitational wave radiation starts to dominate, leading to a merger.
Several studies have shown that stellar dynamical interactions are able to drive the binary to this final stage of evolution and therefore to a merge on a timescale $\lesssim 1$ Gyr  \citep{khan2011,preto2011,vasiliev2015}. 

It is well known that the evolution of the binary eccentricity  after its formation is heavily dependent on the kinematical properties of the surrounding distribution of stars \citep[e.g.,][]{2002ApJ...568..998M}. For example, when most stars are counter-rotating with the SMBHs, the binary eccentricity is expected to increase during the hardening phase,  while the binary eccentricity decreases in co-rotating distributions \citep{iwasawa2011,sesana2011,holleykhan2015,rasskazov2017}. Recent numerical simulations have also shown that an eccentric  binary that is initially retrograde will  reverse its sense of rotation (i.e., flip its orbital plane) roughly at the time of binary formation \citep{mirza2017,khan2019},
from retrograde to prograde. 
Two distinct processes described in the literature can cause angular momentum flips: local 3-body interaction between the binary SMBH and single stars at small separations which reorients the orbital plane \citep[e.g.][]{gualandris2012}, the cumulative effect of dynamical friction due to the rotating environment at larger scales \citep[e.g.][]{dotti2006,dotti2007,bonetti2020}. It is important to note that these two distinct processes only take place for initially retrograde binaries, forming prograde systems.
The binary  becomes co-rotating with respect to the surrounding stellar distribution and the following evolution drives its eccentricity to smaller values. These previous results might therefore suggest  that  SMBH binaries  will have negligible eccentricities when they enter the LISA gravitational wave frequency band.

 In the present paper, detailed $N$-body simulations are used to follow the evolution of merging  galaxies containing a central massive object. We study the evolution of the SMBH binary eccentricity and orbital plane in these models. We show that all binaries are initially prograde with respect to the galaxy sense of rotation. But, the most eccentric binaries ($e\gtrsim 0.7$) rapidly flip their orbital plane reversing their sense of rotation at the moment of binary formation. Thus, contrary to previous work, we find that initially co-rotating binaries become almost perfectly retrograde,  with  their following evolution proceeding towards higher eccentricities. Our models therefore imply  the existence of an eccentric SMBH binary population in LISA. Moreover, we investigate the origin of the orbital flip of the binary using a simple analytical model to describe the early stages of binary evolution. We find that the orbital plane flip is caused by the torque from the surrounding triaxial stellar distribution that characterises  the galactic merger remnant, while excluding that the flip is related to the sense of rotation of stars in the galaxy.

In section \ref{sec:setup} we describe the numerical setup of our $N$-body simulations and method. In section \ref{sec:results} we discuss the results from the models where a flip of the binary orbit is observed. In section \ref{sec:model} we present an analytical model which describes the evolution of a bound binary in a triaxial merger remnant. In section \ref{sec:pop_syn} we present a population synthesis study investigating the flip mechanism and the effect of the galaxy figure rotation. Finally, in section \ref{sec:conclusion} we present our conclusions.

\section{Numerical setup}
\label{sec:setup}

\begin{table}
\begin{tabular}{ccccccc}
\hline 
Simulation & $\gamma$ & $e_i$ & $M_1 : M_2$ & $M_{\bullet}/M_{*}$  & $R/r_0$ & $N$ \\
\hline 
SC05 & 0.5 & 0.5 & 1:1 & 0.005 & 20 & 512k \\
SC07 & 0.5 & 0.7 & 1:1 & 0.005 & 20 & 512k \\
SC09 & 0.5 & 0.9 & 1:1 & 0.005 & 20 & 512k\\
MC05 & 1.0 & 0.5 & 1:1 & 0.005 & 20 & 512k \\
MC07 & 1.0 & 0.7 & 1:1 & 0.005 & 20 & 512k\\
MC09 & 1.0 & 0.9 & 1:1 & 0.005 & 20 & 512k\\
\hline
\hline
\end{tabular}
\caption{Initial parameters of the galaxy merger simulations. From left to right: Simulation identifier; inner slope of the galaxy density profile $\gamma$; initial orbital eccentricity of the progenitor galaxies $e$; mass ratio between the galaxies; SMBH to stellar mass ratio; initial distance between the centres of the two galaxies $R$; total number of particles in the merger simulation $N$.} 
\label{tab:init}
\end{table}

We model mergers of two equal mass galaxies hosting a central SMBH. Each galaxy is described by an isotropic, spherically symmetric \citep{dehnen1993} density profile which is representative of a nuclear bulge
\begin{equation}
\label{eqn:dehnen_density}
	\rho(r) = \frac{(3-\gamma)M}{4\pi} \frac{r_0}{r^{\gamma} (r+r_0)^{4-\gamma}}
\end{equation} 
with inner slope $\gamma$, scale radius $r_0$ and total mass $M$. We choose units such that $G=M_{\rm tot}=r_0=1$, where $M_{\rm tot}$ denotes the total stellar mass of both progenitor galaxies.
We consider two density profiles $\gamma=0.5$ and 1, which represent a shallow and mild cusp and are denoted by SC and MC respectively. The SMBH mass is fixed in all simulations $M_{\bullet} = 0.0025$ yielding a star to SMBH mass ratio of approximately $8\times10^{-4}$. For all the merger simulations considered the progenitor galaxies are placed at an initial distance $R=20r_0$ and on bound elliptical orbits having eccentricities $e=0.5, 0.7, 0.9$. 
All merger simulations were run at a fixed resolution of $N=512\rm{k}$ and lie in the $x$-$y$ plane. 
We also use simulation data from  \citet{Nasim2020}. Specifically, we use the  the Dehnen bulge models with SMBHs, with  $\gamma=0.5$ and $e=0.9$ ($N=512k$ and $N=2048k$, see Table~\ref{tab:other_sims}).
The simulation parameters are given in Table~\ref{tab:init}.

We evolve all merger simulation models with \textsc{griffin} \citep{dehnen2014griffin,Nasim2020,nasim2020_2}, which has recently been shown to evolve SMBH binaries in galaxy merger simulations as accurately as direct summation methods \citep{Nasim2020}.
\textsc{griffin} utilises the Fast Multiple Method (FMM) as a force solver for star-star gravity, avoiding a tail of large force errors, with mean relative force error of $3\times10^{-4}$ (default \textsc{griffin} setting).
We adopt a softening length of $\epsilon_* = 2.3 \times 10^{-2}$ for the stars and $\epsilon_{\bullet}= \epsilon_*/100 = 2.3 \times 10^{-4}$ for the SMBHs and SMBH-star interactions \citep{Nasim2020}.

\section{N-body simulation results}
\label{sec:results}

\subsection{Black hole binary evolution}
\label{subsec:binary evolution}

\begin{figure*}
    \centering
    \includegraphics[width=2.1\columnwidth]{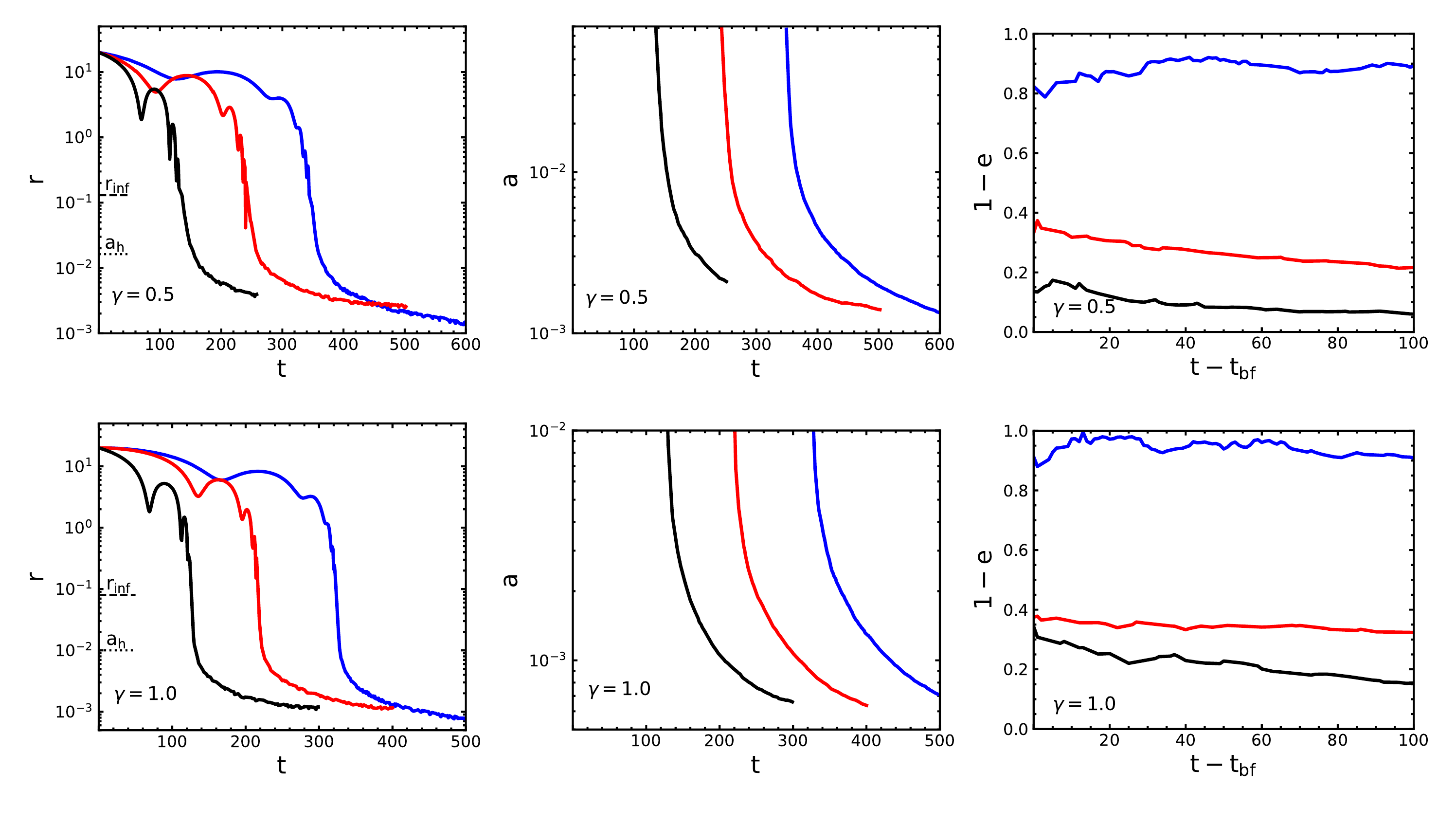}
    \caption{Binary evolution as a function of time for the SC (upper panels) and MC (lower panels) models. From left to right: distance between the SMBHs; semi-major axis evolution and eccentricity evolution from the time of binary formation $\rm{t_{bf}}$. The relevant separations $\rinf$, roughly corresponding to the influence radius of the primary black hole, and the hard-binary separation $\ah$, are marked in the left panels.}
    \label{fig:orbital_elements}
\end{figure*}

In Fig.\ref{fig:orbital_elements} we show the three characteristic evolutionary phases of the binary evolution \citep[e.g.][]{begelman1980,merritt2006,khan2011} for the SC and MC models (see left panels in Fig.~\ref{fig:orbital_elements}).

In the first phase of the merger the orbit of the SMBHs shrinks due to dynamical friction \citep{chandrasekhar1943}. This phase brings the SMBHs to a separation 
at which they form a bound pair. 
{ Here we adopt the convention that a binary is formed when its orbital energy becomes negative.  The orbital energy is defined as
\begin{equation}
 E=-\frac{M_{\rm bin}} {r}+\frac{v^2} {2} ,
\end{equation}
with $M_{\rm bin}$ the binary mass and $v$ the SMBHs relative velocity.} 
Soon after  the binary becomes bound, three body scatterings of stellar particles start also to become important in exchanging energy and angular momentum with the binary.  These encounters result in stellar ejections that lead to a decrease in the central density and the formation of a central core.

As the binary orbit shrinks and becomes more tightly bound, the efficacy of dynamical friction in driving the evolution of the binary decreases and three body interactions start to dominate the evolution. This hardening process takes place rapidly for equal mass binaries, resulting in a swift decrease in the separation until the SMBHs reach the hard binary separation, $\ah$. This is defined as the separation where the binding energy per unit mass exceeds the kinetic energy per unit mass of the stars and can be written as \citep{merritbook2013}
\begin{equation}
\ah = \frac{G\mu}{4\sigma^2} 
\end{equation}
where $\sigma$ is the stellar velocity dispersion and $\mu$ is the reduced mass of the binary ($\mu = M_1M_2/M_{{\rm bin}}$).
A more convenient definition of the hard binary separation, as $\sigma$ is not constant, that we adopt below is given by
\begin{equation}
    \ah \equiv \frac{\mu}{M_{\rm bin}}\frac{r_{\rm inf}}{4} = \frac{r_{\rm inf}}{16}, 
\label{eq: hard_binary_sep}
\end{equation}

where $r_{\rm inf}$ is the influence radius of the primary SMBH, defined as the separation at which the stellar mass within the orbit of the binary $M_*$ is equal to twice the mass of the primary black hole $M_\bullet$, i.e.,
\begin{equation}
    M_*(<r_{\rm inf}) = 2M_{\bullet} \ .
\label{eq: ah}
\end{equation}

In the left panels of Fig.~\ref{fig:orbital_elements}
we illustrate the evolution of the binary separation and the values of the characteristic separations $\rinf$ and $\ah$ for our  models.
We can see that the evolution of the SMBHs is  dependent on the density profile and the initial orbital eccentricity, with the denser models (MC) with the larger eccentricities evolving more rapidly. 
{In the middle panels of Fig.\ref{fig:orbital_elements}
we plot the evolution of the semi-major axis after the orbital energy of the binary becomes positive which sets the time of binary formation.
In the SC models the binary forms with a semi-major axis  $a\simeq 0.08$, while in the MC models it forms with $a\simeq 0.01$.}
In the right  panels of Fig.\ref{fig:orbital_elements}, we observe a significant variation in the binary eccentricity at the moment of binary formation  among both SC and MC models, with  the binaries forming with larger eccentricities in the more eccentric galaxy mergers.
This apparent correlation showing that more eccentric mergers generally result in a larger binary eccentricity has been observed in previous studies \citep[e.g.][]{khan2011}, though it is important to note that at the resolution of these studies stochastic effects play a significant role in the eccentricity variation \citep{Nasim2020}.
Somewhat unsurprisingly, we find that the most eccentric mergers ($e_i=0.9$) yield the most eccentric binaries for both the SC and MC models reaching $e\gtrsim 0.95$ and $e\gtrsim 0.90$ respectively by the end of the numerical integration (see Table \ref{tab:flip_table}). 

\subsection{Binary orbital plane flip} \label{bopf}

\begin{table}
\centering
\caption{ Binary  parameters in our merger models. From left to right: Simulation identifier; final eccentricity of the binary at the end of the numerical integration $e_{\rm{final}}$; time at which we observe the flip of the orbital plane of the binary  $\rm{t_{flip}}$ ; time of binary formation $\rm{t_{bf}}$ defined as the time at which the binary separation becomes less than $\rinf$. }
\begin{tabular}{cccc}
\hline 
Models & $e_{\rm{final}}$ &  $\rm{t_{flip}}$ & $\rm{t_{bf}}$ \\
\hline 
SC05 &  0.10 & N/A & 350 \\
SC07  &  0.80 & N/A & 247 \\
SC09  &  0.95 & 128 & 129 \\
MC05  &  0.11 & N/A & 340  \\
MC07  &  0.71 & N/A & 233 \\
MC09  &  0.91 & 123  & 125 \\
\hline
\end{tabular}
\label{tab:flip_table}
\end{table}

\begin{figure*}
    \centering
    \includegraphics[width=1.8\columnwidth]{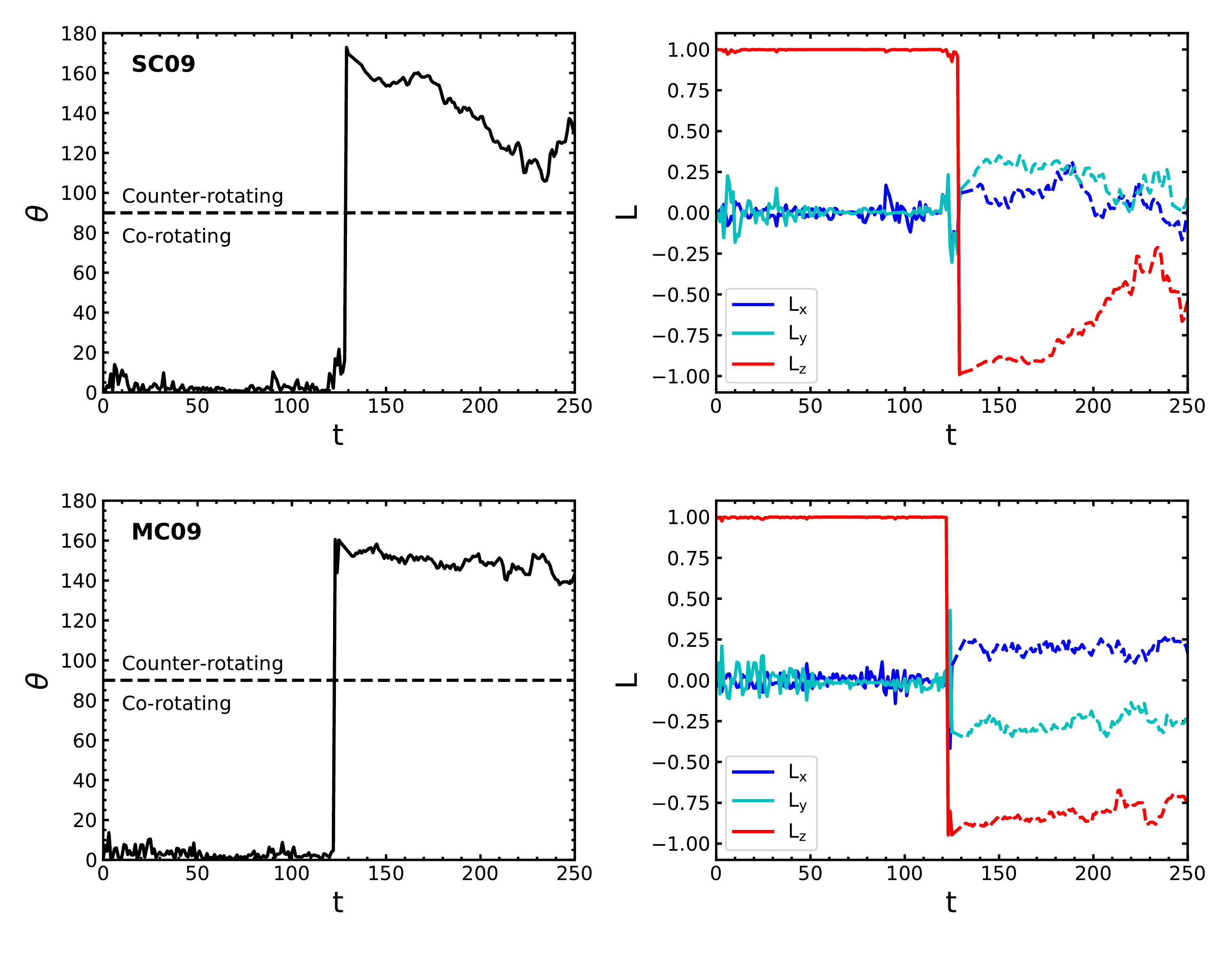}
    \caption{Evolution of the angle between the binary and galaxy angular momenta vectors ($\theta$), and the normalised angular momentum components of the binary ($\rm{L_{x}, L_{y}, L_{z}}$) for the models where we observe the binary orbital plane flip (SC09 and MC09, upper and lower panels respectively). The black horizontal dashed line in the left panels represent the boundary between co-rotating and counter-rotating binaries. The solid/dashed lines in the upper right panel represent the angular momentum components before/after binary orbital plane flip. We observe that the binary orbital plane itself flips, causing the transition from co-rotating to a counter-rotating with respect to the galaxy angular momenta.}
    \label{fig:theta_angmom}
\end{figure*}

 We consider here the evolution of the angle, $\theta=\cos^{-1}\left( \bm{L}\cdot \bm{L}_{\rm gx}/|\bm{L}||\bm{L}_{\rm gx}| \right)$, between the binary   angular momentum vector, $\bm{L}$, and the galaxy total angular momentum vector, $\bm{L}_{\rm gx}$. We compute the angular momentum of the binary using $\bm{L} = \bm{r} \times \bm{v}$, where $\bm{r}$ and $\bm{v}$ are the relative position and velocity between the SMBHs. We compute the angular momentum of the galaxy $\bm{L}_{\rm gx}$ relative to the centre of mass of the merging progenitors which coincides with the centre of mass of the binary.

 In all our models, the binary is initially co-rotating with the galaxy and $\theta\sim 0$. 
 The subsequent evolution of  $\theta$  is shown in Fig.\ref{fig:theta_angmom} for models SC09 and MC09, together with the normalised angular momentum components of the binary. We observe that in these models the binary  switches from co-rotating to counter-rotating. The flip occurs over a time interval $\Delta t < 1$ approximately at the time of binary formation (see Table \ref{tab:flip_table}). At the same time,  we see that there is a significant change in the binary angular momentum component $\rm{L_{z}}$, which  changes from positive to negative. These characteristics show that the binary orbital plane itself is flipping along the $z$-axis at the time the binary becomes counter-rotating. Importantly, we only observe this behaviour in the most eccentric models (SC09 and MC09), with the other models simply resulting in a co-rotating binary. 
 We find that flips occur for eccentric binaries where the binary eccentricity is initially $e\lesssim 0.7$ the binary remains co-rotating until the end of the simulation. On the other hand, in models where $e\gtrsim 0.7$, the binary reverses its sense of rotation and becomes almost perfectly retrograde at the time of binary formation. 
 
 \begin{table}
\begin{tabular}{cccccccc}
\hline 
Suite & $\gamma$ & $e_i$ & $M_1 : M_2$ & $M_{\bullet}/M_{*}$  & $R/r_0$ & $N$ & $N_r$ \\
\hline 
MR & 0.5 & 0.9 & 1:1 & 0.005 & 20 & 512k & $8$\\
HR & 0.5 & 0.9 & 1:1 & 0.005 & 20 & 2048k & $4$\\
\hline
\end{tabular}
\caption{Initial parameters of the galaxy merger simulations from \citet{Nasim2020}. From left to right: Simulation suite identifier; inner slope of the galaxy density profile $\gamma$; initial orbital eccentricity of the progenitor galaxies $e$; mass ratio between the galaxies; SMBH to stellar mass ratio; initial distance between the centres of the two galaxies $R$; total number of particles in the merger simulation $N$; number of random realisations $N_r$.} 
\label{tab:other_sims}
\end{table}

\begin{figure}
    \centering
    \includegraphics[width=1.\columnwidth]{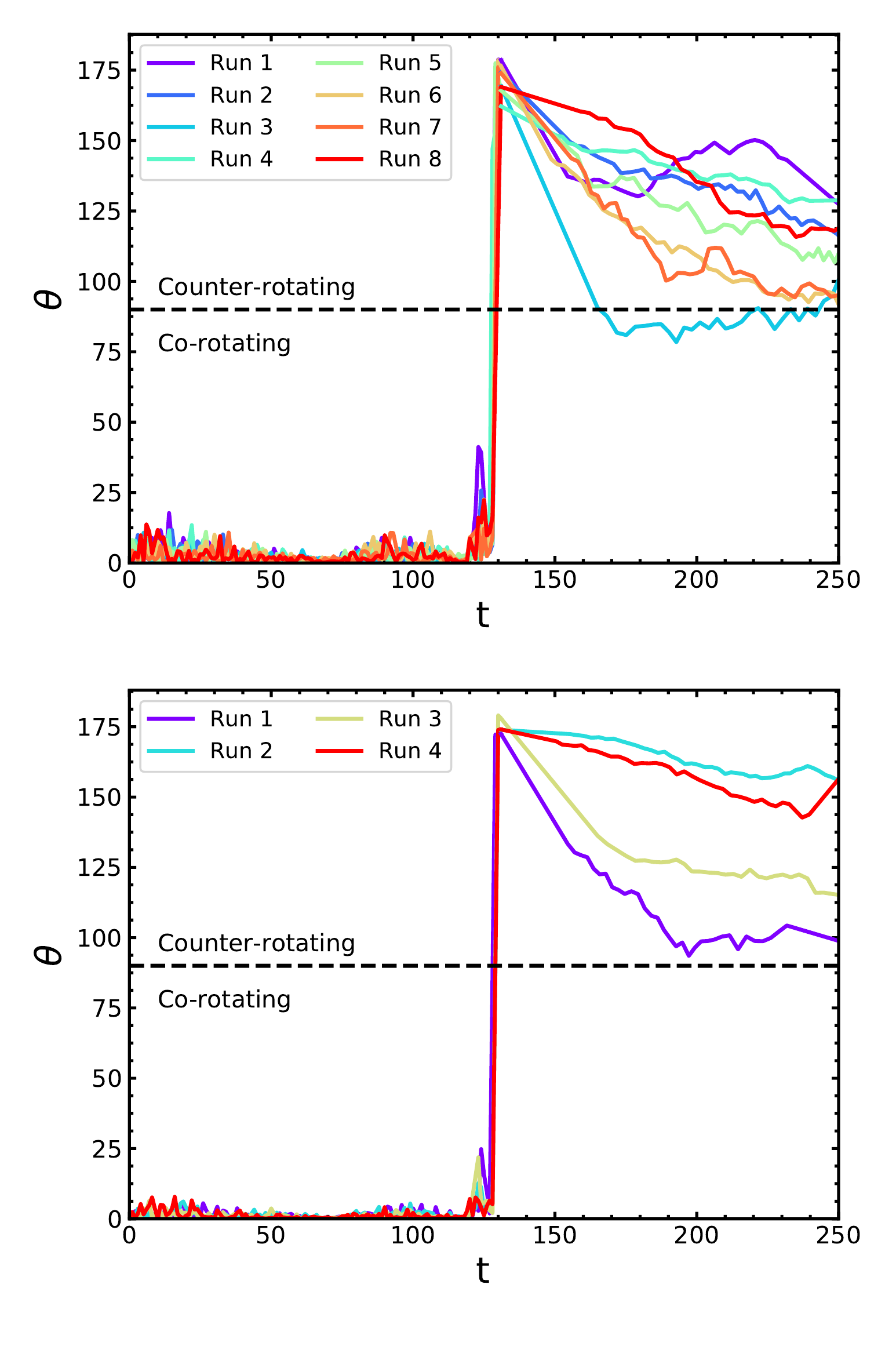}
    \caption{Evolution of the angle between the binary and galaxy angular momenta vectors ($\theta$) for the different realisations with SC09 parameters. The top panels shows the results for the 8 different realisations at $N=512k$ and the bottom panels shows the 4 different realisations at $N=2048k$. We observe a binary orbital flip for all realisation irrespective of the numerical resolution. }
    \label{fig:theta_realisations}
\end{figure}

To check that the behaviour we observe is not due to our specific generation of initial conditions or  to numerical resolution (i.e. not a spurious result), we use the simulation data from \citet{Nasim2020} who considered 8 realisations with the SC09 models parameters at $N=512k$ and 4 realisations at $N=2048k$. 
The parameters for these numerical simulations are presented in Table~\ref{tab:other_sims}.
We plot the evolution of $\theta$  for the different realisations in Fig.\ref{fig:theta_realisations}. For all of the realisations, irrespective of resolution, we observe that 
the binaries show the same behaviour, becoming counter-rotating at about the same time. 
After the binaries flip we see that $\theta$ is slightly different between realisations. This is expected due to the stochastic nature of the interaction between the binary and the surrounding stellar environment \citep[e.g.][]{bonetti2016,bortolas2018,Nasim2020}. From Fig.\ref{fig:theta_realisations} we observe a clear trend at later times: after the binary flips the value of $\theta$ systematically decreases for all realisations. This is due to the fact that misaligned binaries tend to align their angular momentum vector with the angular momentum vector of the surrounding stellar system on a characteristic time-scale of a few hardening times \citep{gualandris2012}. It was shown that this process is linear with time and is independent on the mass ratio between the stellar particles and the binary members. These theoretical expectations are in  agreement with our numerical results.

\begin{figure}
    \centering
    \includegraphics[width=1.\columnwidth]{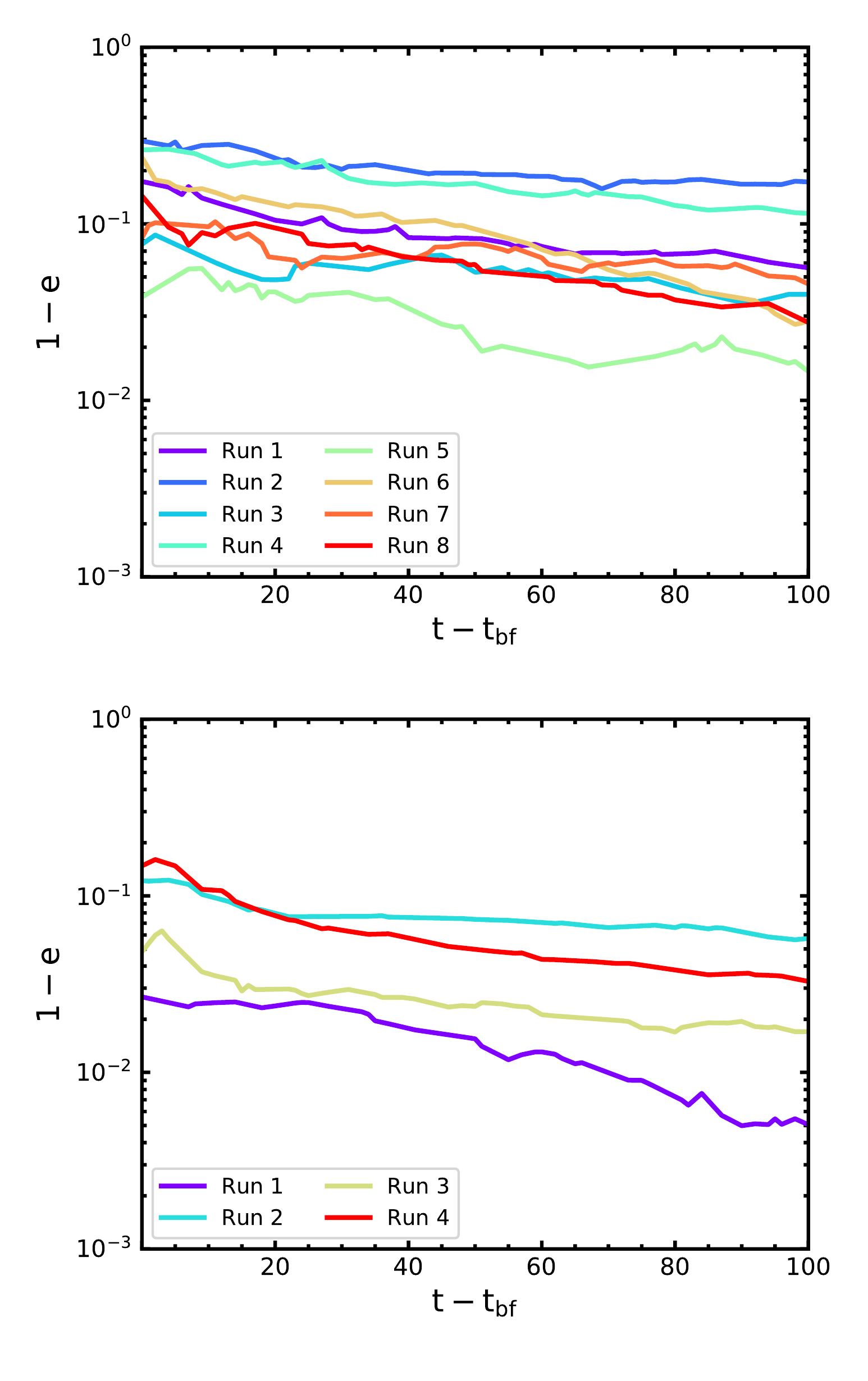}
    \caption{Evolution of the eccentricity after the time of binary formation ($\rm{t_{bf}}$) for the different realisations of SC09. The top panels shows the results for the 8 different realisations at $N=512k$ and the bottom panels shows the 4 different realisations at $N=2048k$. For all realisations we observe an increase in the eccentricity after the time of binary formation, with the majority of binaries reaching significantly large eccentricities ($e \gtrsim 0.9$). }
    \label{fig:ecc_realisations}
\end{figure}

In Fig.\ref{fig:ecc_realisations}, we plot the eccentricity evolution after the time of binary formation for the different realisations of SC09. We observe that the eccentricity increases in all binaries, enabling them to achieve large eccentricities ($e \gtrsim 0.9$) by the end of the integration. For the higher resolution case (see Fig.\ref{fig:ecc_realisations}, bottom panel), we observe that the more eccentric binaries are characterised by a faster decrease in  $\theta$, with the binary reaching an extremely large eccentricity by the end of the evolution ($e \gtrsim 0.995$).

A similar orbital plane flip has been observed in previous studies \citep[e.g][]{mirza2017,khan2019}.
However, we note some key differences. \citet{mirza2017} and \citet{khan2019} consider the evolution of SMBH binaries in equilibrium axisymmetric galaxy  models; they introduce a net stream rotation by flipping the direction of the
velocity components of the $N$-body particles in their models. They show that during the
pairing phase, the eccentricity of the binary dramatically increases for retrograde configurations and the orbital plane flips so that the binary becomes prograde,
which suppresses the further eccentricity growth. They find that in initially prograde configurations, SMBH binaries form and remain prograde.
Instead, our models are formed by galaxy merger simulations and are therefore triaxial systems \citep[e.g.,][]{bortolas2018} and we find that the orbital plane flip takes place in an initially co-rotating stellar distribution leading to a counter-rotating binary, which has not been explicitly shown before.
We believe, however, that the flip found in earlier work and the  behaviour described in this paper might be related. We will come back to this point in Section \ref{sec:pop_syn}.

There are at least three mechanisms that could lead to the reorientation of a massive binary in a galactic nucleus: (i) repeated
close interactions with passing stars  \citep[e.g.,][]{2002ApJ...568..998M,gualandris2012,rasskazov2017};
(ii) the cumulative dynamical friction due to the rotating environment at larger scales \citep{dotti2006,bonetti2020};
and (iii) the torque due to the non-axisymmetric stellar potential around the binary, the mechanism  proposed in this paper.

In spherical and isotropic cusps, due to the lack
of any preferential direction in these systems, the orbital
plane of the binary can only undergo a random walk, resulting in small changes in the orbital plane on long timescales \citep[e.g.,][]{2002ApJ...568..998M}. 
However, 
in rotating stellar systems a
significant reorientation is expected \citep{gualandris2012,rasskazov2017}.
Binaries whose angular momentum is initially misaligned with respect to that of the stellar cusp tend to realign
their orbital planes with the angular momentum of the cusp on a timescale of a few
hardening times. Since
 the SMBH binaries in our models start with their angular momentum  aligned with that of the stellar cusp, 
this process is not expected to significantly affect their orientation. Thus, we conclude that mechanism (i) cannot be responsible for the flip observed in the $N$-body models.

\begin{figure}
    \centering
    \includegraphics[width=0.8\columnwidth]{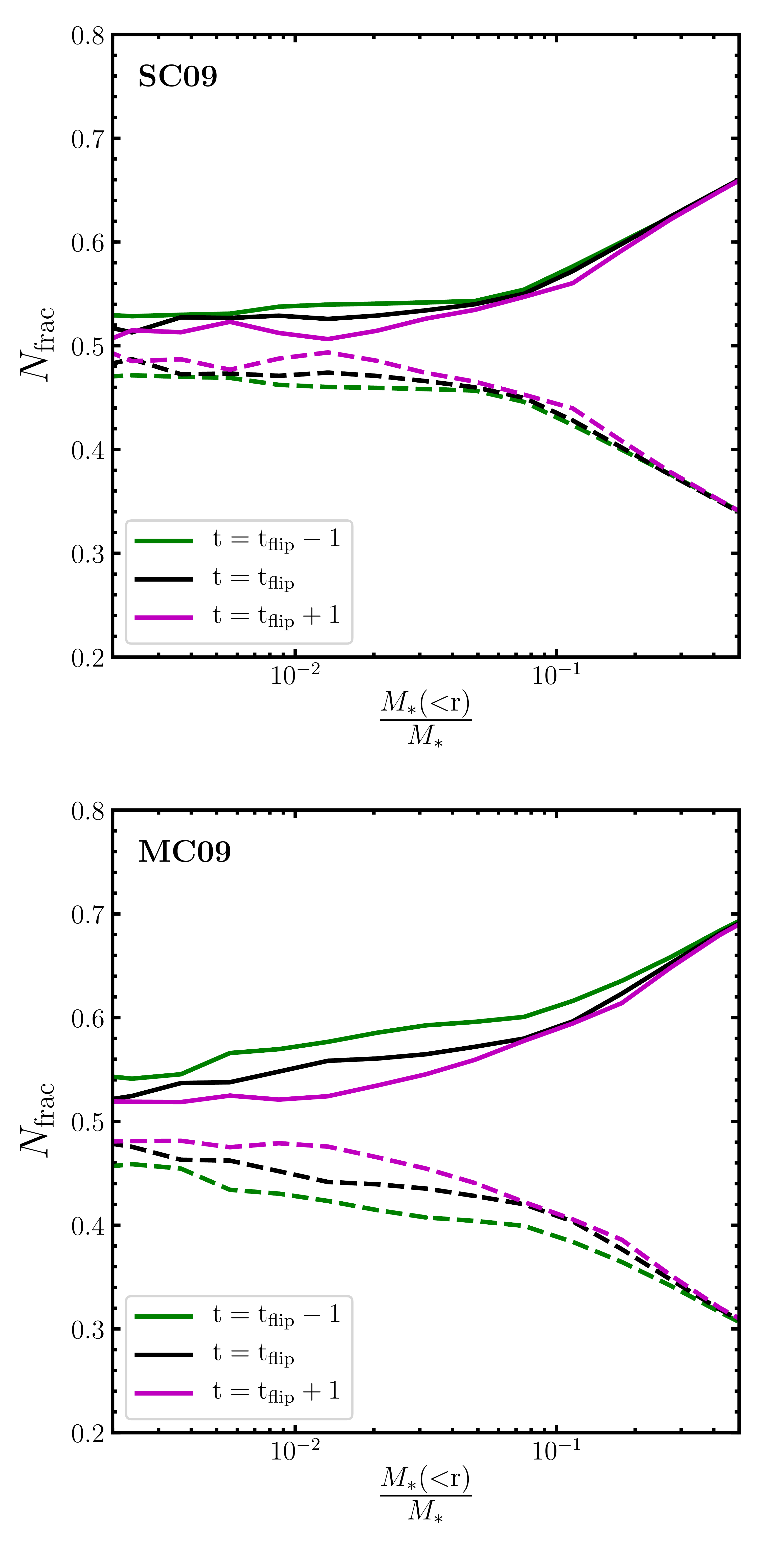}
    \caption{Fraction of stars on retrograde (dashed line) and prograde (solid line) orbits, with respect to the angular momentum vector of the binary \textit{prior} to flip, as a function of the normalised enclosed stellar mass. The coloured lines are defined as: time unit before the binary flip (green line), first time where we observe the flip (black line), subsequent time unit after the flip has been observed (magenta line). Upper/lower panels show the results from models SC09/MC09. We do not observe any significant evolution in the prograde/retrograde stellar fraction before and after the flip indicating the flip mechanism is an independent process.}
    \label{fig:pro_retro_frac_flip}
\end{figure}

From our results we also check whether there is a dynamical effect related to the ejection of stars present in the galaxy. To do this, we plot in Fig.\ref{fig:pro_retro_frac_flip} the fraction of prograde and retrograde stars with respect to the angular momentum vector of the binary \textit{prior} to flipping and as a function of the enclosed mass at times before and after the flip takes place. We observe no significant change in the fraction of prograde $vs$ retrograde orbits before and after the flip takes place. This shows that the mechanism causing the flip is independent of the sense of rotation of the stars in the galaxy, and supports our previous statement that  the binary flip is not caused by
 the angular momentum exchange between stars and the SMBH binary during close encounters, a mechanism discussed by \citet{sesana2011} and \citet{gualandris2012}.
Such encounters with stars are  responsible, however,  for the evolution of the binary at later times ($t\gtrsim130$ in Fig.\ref{fig:theta_angmom}), which tend to reorient the system towards a co-rotating state.

The process proposed by \citet{dotti2006} posits that dynamical friction causes the angular momentum flip, whereby a massive perturber is accelerated in the opposing direction to its instantaneous motion relative to the local environment. As dynamical friction always decelerates counter-rotating systems, their eccentricity proceeds to increase (without flipping the orbital plane) until the effect over one orbit is sufficient enough to reverse the sense of rotation, causing a gradual angular momentum flip. Recently, numerical tests presented in \citet{bonetti2020} confirmed that dynamical friction was responsible for the flip observed in \citet{dotti2006}, and dubbed the mechanism ``drag toward circular corotation". 
This model also suggests that massive perturbers that are initially co-rotating will circularise their orbit and  remain prograde.
We conclude that mechanism (ii) also cannot be responsible for the flip  we observe.

The behaviour resulting in the formation of counter-rotating binaries suggests that there is a connection between the orbital plane flip and the binary eccentricity, with very eccentric mergers yielding an orbital flip. 
Interestingly, it has been shown that more eccentric galactic mergers yield a larger level of triaxiality in the merger remnant \citep[e.g.][]{bortolas2018}. To verify this, we determine the triaxiality of the merger remnant for models in which a flip is observed.

The shape of a cluster can be characterised by the best fitting ellipsoid to the stellar distribution at an arbitrary distance. If $a>b>c>0$ are the axes of symmetry of this ellipsoid, the triaxiality can be defined as
\begin{equation}
T=\frac{a^2-b^2}{a^2-c^2}.
\end{equation}
where if $0\leq T <0.5$ the spheroid is oblate, if $T=0.5$ the system is said to have maximum triaxiality, and if $0.5 < T \leq 1$ the spheroid is prolate.
To compute the model's axes 
we use an iterative method similar to that described in \citet{katz1991} \citep[see also][]{antonini2009,bortolas2018}. Briefly, we 
initially consider all  stars enclosed within a sphere of radius equal the radius of influence of the binary \footnote{We find that we obtain the same triaxiality by considering distances equal to two or three times the radius of influence of the binary. This confirms that the computation of the triaxiality at different distances does not affect our measure of triaxiality.}, and 
 compute the inertia tensor; from this we then compute the eigenvalues of the inertia tensor which serves as a first order approximation for the axes of symmetry, and the three direction cosines giving the direction of the
eigenvectors. To achieve a greater level of accuracy we iterated the procedure by computing new axes, but now only considering particles enclosed within the ellipsoid with the previously computed axes. This procedure is continued until the tolerance ($10^{-5}$) is satisfied.

Computing the triaxiality parameter of models SC09 and MC09 at the point where the binary flips yields $T=0.53$ and $T=0.56$ respectively, with axis ratios $b/a$, $c/a$ of [0.85, 0.69] and [0.84, 0.69].
These levels of triaxiality are in agreement with the results found by \citet{bortolas2018}. We verified that these values are independent of the numerical resolution  by computing the triaxiality for the different realisations of model SC09, which resulted in similar values of $T$. These levels of triaxiality are extremely large, corresponding  almost to maximum triaxiality. The less eccentric galactic mergers produce a less triaxial merger remnant, with SC05 and SC07 having $T=0.71, 0.68$, while models MC05 and MC07 have triaxiality $T=0.73, 0.69$ respectively.
In the previous analysis, we classified a binary ``flip" with respect to the angular momentum of the galaxy. However, we find that $\bm{L}_{\rm gx}$ is aligned within
$2\deg$ with  the shortest axis  of our best fitting ellipsoid; in our setup, this direction also coincides  with the $z$ axis. In the examined scenario, the merger itself determines the rotation/minor axis orientation, making small inclinations more probable.
Thus, one can alternatively state that  the binary flip in our models occurs with respect to the minor axis of the galaxy. As argued below,  this latter statement turns out to give, in fact, a more correct representation of the binary evolution seen in the $N$-body models.

The large triaxiality of the models, especially in the most eccentric models, and the high initial eccentricity of the binaries suggest that the orbital flip observed in the $N$-body models is driven by the long-range tidal torques arising from the surrounding galaxy. We investigate this possibility in the following sections.
We are not fully certain on the reason why the flips take place approximately at the time of binary formation.

\section{Semi-Analytical model}
\label{sec:model}
In this section we use a simple analytical model to investigate the possibility that the sudden reorientation of the binary orbital plane observed in the  $N$-body models is a consequence of the  torque from the triaxial background potential. 
We stress that the flip of the SMBH binary orbital plane in our $N$-body models occurs in less than $\sim1$ $N$-body time units at the moment of binary formation.  The semi-analytical model presented here is therefore intended to describe the evolution of the SMBH just after this time.

We consider a bound binary system of total mass $M_\mathrm{bin}$ with semi-major axis $a$ embedded within a merger remnant with a density distribution $\rho({\bf r})$. 
We assume that the orbit of the binary is Keplerian and model it using the vectorial formalism  \citep[e.g.,][]{tremaine2009, tremaine2014}. Thus, we define the vectors
\begin{equation}
\label{eq:base_vectors}
    {\bf{e}} \equiv e \,{\bf \hat{e}} \hspace{2em} {\bf{j}} \equiv \sqrt{1 - e^2} \, {\bf \hat{j}}
\end{equation}
where $\bf{\hat{j}}$ is parallel to the angular momentum vector, $\bf{\hat{e}}$ points toward the pericenter of the binary and $e$ is the eccentricity. We define Cartesian unit vectors $\hat{\bf n}_x$, $\hat{\bf n}_y$, and $\hat{\bf n}_z$, which provide the reference frame with respect to the centre of mass  of the density distribution.
For simplicity, we assume that the centre of mass of the binary system lies at the centre of the density distribution.

\subsection{Basic equations} \label{equa}
Following \citet{PA2017},  we model the triaxial potential of the merger remnant per unit mass as
\begin{equation}\label{eq:triaxial_potential}
\begin{split}
{\Phi}({\bf r}_i)&=\frac{4\pi G}{(3-\gamma)(2-\gamma)}
\tilde{\rho} \tilde{r}^2\left(\frac{r_i}{\tilde{r}}\right)^{2-\gamma}\\
& \times\left[1+
\epsilon_z\frac{(\hat{\bf n}_z\cdot {\bf r}_i)^2}{r_i^2}+
\epsilon_y\frac{(\hat{\bf n}_y\cdot {\bf r}_i)^2}{r_i^2}\right],
\end{split}
\end{equation}
with $0<\epsy \lesssim \epsz$ and ${\bf r}_i$ represents the position of the BH $i=1,2$ relative to the remnant's origin. Assuming that the BHs have equal masses and their center of mass is at the origin such that their relative position is $\Delta {\bf r}={\bf r}_1-{\bf r}_2=2{\bf r}_1$, we can express the total potential for the binary pair as
\begin{equation}
\label{eq:phi_bin}
{\Phi}_{\rm bin}\equiv{\Phi}({\bf r}_1)+{\Phi}({\bf r}_2)=2{\Phi}[\Delta {\bf r}/2]=2^{\gamma-1}{\Phi}(\Delta {\bf r}).
\end{equation}
This potential (${\Phi}\propto{\Phi}_{\rm bin}$) has the advantage that it allows for an analytical treatment of the equations of motion.
The density distribution associated with this triaxial potential can be obtained by solving the Poisson equation  and is given by \citep[e.g.][]{merritbook2013}
\begin{eqnarray}
\rho({\bf r})=\tilde{\rho}\left(\frac{r}{\tilde{r}}\right)^{-\gamma}
\left[1+\tilde{\epsilon}-
\tilde{\epsilon}_z\frac{({\bf r}\cdot \hat{\bf n}_z)^2}{r^2}
-\tilde{\epsilon}_y\frac{({\bf r}\cdot \hat{\bf n}_y)^2}{r^2}\right],\nonumber\\
\label{eq:rho_eps}
\end{eqnarray}
where
\begin{eqnarray}
\tilde{\epsilon}_i=\epsilon_i\frac{\gamma(5-\gamma)}{(2-\gamma)(3-\gamma)},
\end{eqnarray}
with $i=y,z$ and
\begin{eqnarray}
\tilde{\epsilon}=\frac{2}{(2-\gamma)(3-\gamma)} (\epsilon_y + \epsilon_z).
\end{eqnarray}
For the models considered in this study, we choose $\gamma=1.0$.
In order to compare to the $N$-body models, it is also useful to relate $\epsilon_i$ to the galaxy axis ratios. For $\gamma=1$ we have
\begin{align}\label{arat}
\frac{b}{a}=\frac{1+\tilde{\epsilon}-
\tilde{\epsilon}_y} {1+\tilde{\epsilon}},\
{\rm and}\ 
{\frac{c}{a}}=\frac{1+\tilde{\epsilon}-
\tilde{\epsilon}_z} {1+\tilde{\epsilon}} \ ,
\end{align}
where similar to our definition above, $a$,
$b$, and $c$ represent  the axis length scales
of the  mass distribution along $\hat{\bf n}_x$, $\hat{\bf n}_y$, and $\hat{\bf n}_z$,
respectively. Since the isodensity contours are not ellipsoids, the axis ratios are defined here in terms of the points of intersection of  an isodensity contours with the $y-$, $z-$ and $x-$axes\footnote{This definition is different from that in \citet{PA2017} that uses approximate ellipsoidal isodensity contours, which are only valid for $r\sim r_0$.}.

In the Appendix \ref{ap:equations_of_motion}, we averaged the equations of motion for $\Phi$ with $\gamma=1$ (Eq. \ref{eq:triaxial_potential}), for which $\Phi=\Phi_{\rm bin}$ (Eq. \ref{eq:phi_bin}), resulting in
\begin{align}
&\langle\Phi_{\rm bin}\rangle =\pi G\tilde{\rho} \tilde{r}a 
\left\{3-j_{}^2+\sum_{i=y,z}  \epsilon_i 
\left[j_{}^2+3\left(\hat{\bf n}_i\cdot {\bf {e}}_{}\right)^2
-\left(\hat{\bf n}_i\cdot {\bf {j}}_{}\right)^2 \right]\right\}.
\label{eq:tot_potential}
\end{align}
Thus the orbit evolution of the binary is given by the following equations of motion (equations [\ref{eq:ap_dj_dt}-\ref{eq:ap_de_dt}]):

\begin{align}\label{eq:dj_dt}
\frac{d {\bf {j}}}{d\tau} =  \sum_{i=y,z}  \epsilon_i 
\left[3\left(\hat{\bf n}_i\cdot {\bf {e}}_{}\right) \left({\bf e} \times \hat{\bf n}_i\right)-\left(\hat{\bf n}_i\cdot {\bf {j}}_{}\right) \left({\bf j} \times \hat{\bf n}_i\right)  \right]
\end{align}
\begin{align}\label{eq:de_dt}
\frac{d {\bf {e}}}{d\tau} =\sum_{i=y,z}  \epsilon_i 
\left[3\left(\hat{\bf n}_i\cdot {\bf {e}}_{}\right) \left({\bf j} \times \hat{\bf n}_i\right)-\left(\hat{\bf n}_i\cdot {\bf {j}}_{}\right) \left({\bf e} \times \hat{\bf n}_i\right) \right]
\end{align}
where we have introduced the dimensionless time $\tau=t/\tau_{\rm sec}$ with 
\begin{equation}
    \tau_{\rm{sec}} = \frac{\sqrt{M_{\rm{bin}}}}{2\pi \tilde{r} \tilde{\rho} \sqrt{Ga} }.
    \label{eq:secular_time}
\end{equation}
We note that the functional form of our equations of motion are equivalent to that  in \citet{PA2017} which describes the evolution of the outer orbit in their setup of an inner binary orbiting a SMBH.

Given that the merger remnant is not fully spherically symmetric, the binary angular momentum ${\bf j}$ will precess around both $\hat{\bf n}_z$ and $\hat{\bf n}_y$  with  characteristic timescales given by
\begin{align}
 \tau_{\rm{sec,z}} = \frac{\tau_{\rm{sec}}}{\epsz} = \frac{\sqrt{M_{\rm{bin}}}}{2\pi \epsz \tilde{r} \tilde{\rho} \sqrt{Ga} },
 \label{eq:tau_scz}
 \end {align}
\begin{align}   \label{eq:tau_scy}
  \tau_{\rm{sec,y}} = \frac{\tau_{\rm{sec}}}{\epsy} = \frac{\sqrt{M_{\rm{bin}}}}{2\pi \epsy \tilde{r} \tilde{\rho} \sqrt{Ga} }.
\end{align}

From equations~(\ref{eq:dj_dt}) and~(\ref{eq:de_dt}) 
 we see that the results of our models only depend  on the shape parameters of the  galaxy $\epsilon_i$, the power-law index $\gamma$ and on the final integration time. The results can then be rescaled from the expression of the secular time, $\tau_{\sec}$, with the required values of the parameters $(\tilde{r},\tilde{\rho},a,M_{\rm bin})$.
For example, in order to approximately compare the results of the integrations presented below to the $N$-body models MC, 
we start by assuming that the mass density profile is spherically symmetric so that  $ \rho(r)= \tilde{\rho} (r/\tilde{r})^{-\gamma}$,  which is consistent with the Dehnen profile for $r \ll r_0$. We then set $\tilde{r}=r_{\rm inf}$ so that for $\gamma=1$ one finds $\tilde{\rho}=M_\bullet/(\pi r_{\rm inf}^3$). 
Using equation~(\ref{eq:secular_time}), this leads to  $\tau_{\rm sec}/P_{\rm bin}=(r_{\rm inf}/a)^2/(2\pi)$,
where $P_{\rm bin}$ is the binary orbital period.
In the $N$-body models, $P_{\rm bin}\simeq0.09$, $r_{\rm inf}\simeq 0.1$, and $a\simeq 0.01$  at the moment of binary formation (approximately). Thus, we have $\tau_{\rm sec}/P_{\rm bin}\simeq 16$, or  $\tau_{\rm sec}\simeq 1.4$ in the time units of the $N$-body simulations.

\subsection{Limitations of the model}

Although the equations of motion  (\ref{eq:dj_dt}) and (\ref{eq:de_dt}) allow for a simple and insightful treatment of the general dynamics of a binary in a triaxial field, they come at the expense of making a range of simplifying approximations when compared to the $N$-body simulations.
Firstly, at the moment of binary formation the galaxy mass distribution is still significantly affected by the motion of the binary as it hardens by scattering off surrounding stars. This effect is not included in the analytical models, which assume a fixed density distribution of stars.

Secondly, when carrying out the orbit average (also known as the secular approximation), the semi-major remains constant. This is unlike the evolution depicted in Fig.\ref{fig:orbital_elements} that includes the effect of the binary hardening. However, as shown in our $N$-body simulations (see Fig.\ref{fig:orbital_elements}) the semi-major axis evolves on longer timescales compared to the binary flip timescale and thus a constant semi-major axis is a fair approximation to make. We note here that hardening and dynamical friction could be included in this secular formalism (e.g., \citealt{DA2017}), but it goes beyond the goal of our simplified model that attempts to only capture the basic conditions for orbit flipping driven by the torques from the host galaxy.

Finally, the orbit-averaged procedure assumes that changes in the binary orbit occur on a timescale much longer than the orbital period of the binary. This is partially satisfied since  in our $N$-body simulations  $\tau_{\rm sec}/P_{\rm bin}\simeq 16$ at the moment the binary flip occurs.
To compare directly with our numerical simulations, we only consider equal mass binaries.

\section{Conditions for orbital flip}
\label{sec:pop_syn}

\subsection{Flip fraction}
\label{sec:flip_frac}

\begin{figure*}
    \centering
    \includegraphics[width=2.1\columnwidth]{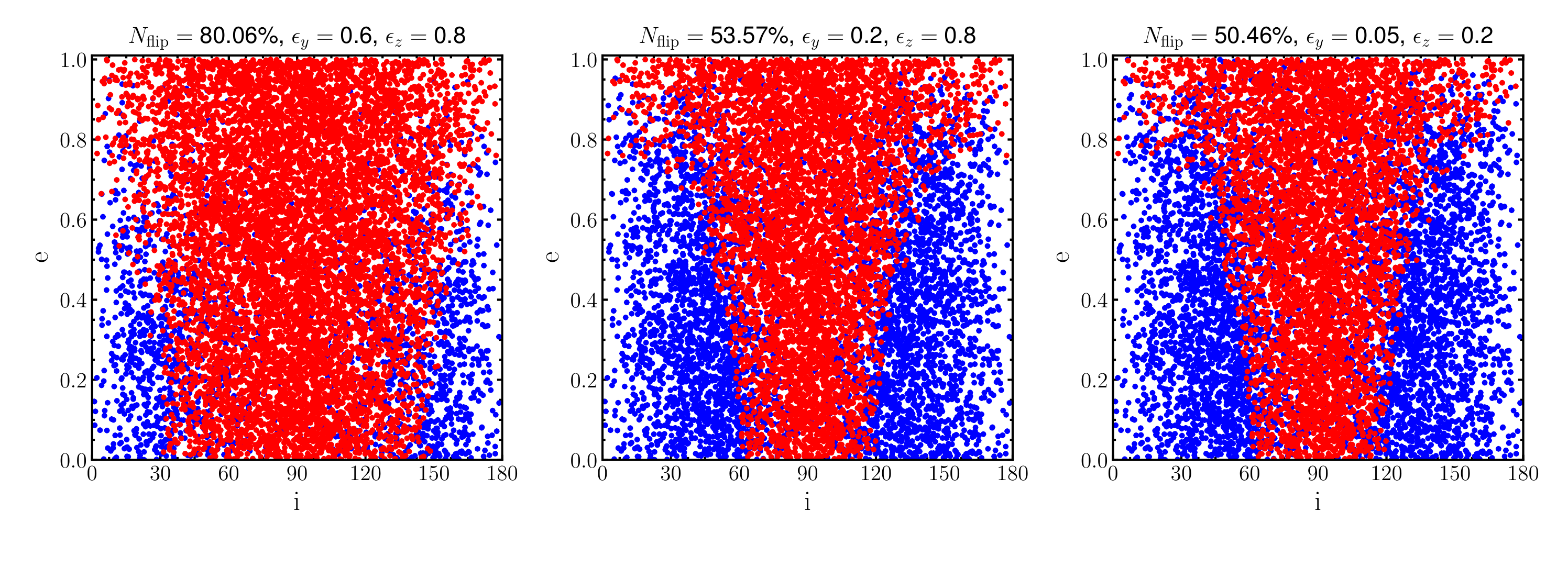} 
     \includegraphics[width=2.15\columnwidth]{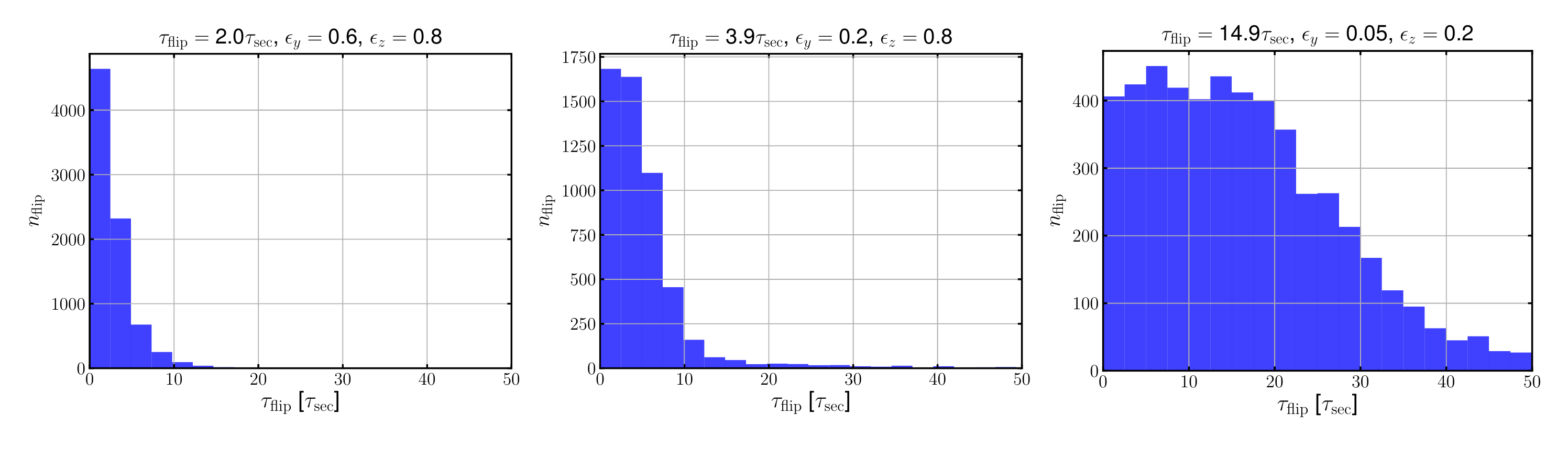} 
    \caption{The initial eccentricity and inclination of the 10000 binaries run in the population synthesis model. All binaries were run for $50$ secular times $\tau_{\rm{sec}}$, the blue filled circles represent the binaries that did not flip with respect to the $\hat{\bf n}_z$ and the red filled circles represent the binaries that flip. The upper left panel shows the flip fraction for different levels of flattening: strong (right panel), intermediate (middle panel), and weak (right panel). The values for $\epsy$ and $\epsz$ are given at the top of each panel together with the flip fraction $N_{\rm{frac}}$. 
    The bottom panels show the distribution of the flip times $\tau_{\rm{flip}}$, with the median flip time given at the top of each panel.
    We clearly observe a significant fraction of binaries that flip, with the potentials with the larger values of $\epsy$ and $\epsz$  leading to a larger flip fraction. 
    We also observe that the flip times are rapid, decreasing for increasing levels of flattening.}
    \label{fig:pop_synth_flip}
\end{figure*}

\begin{figure}
    \centering
    \includegraphics[width=0.85\columnwidth]{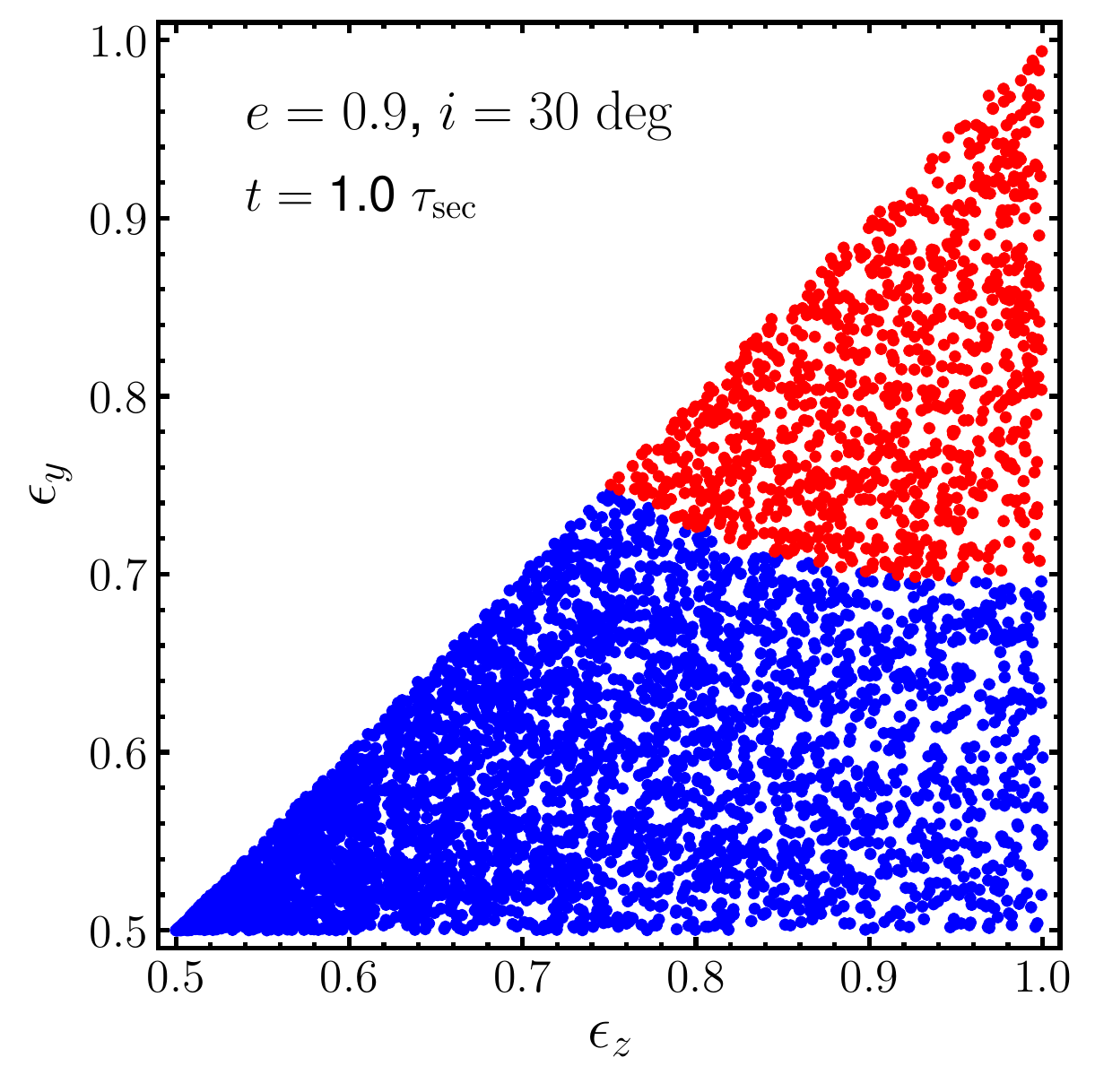} 
     \includegraphics[width=0.85\columnwidth]{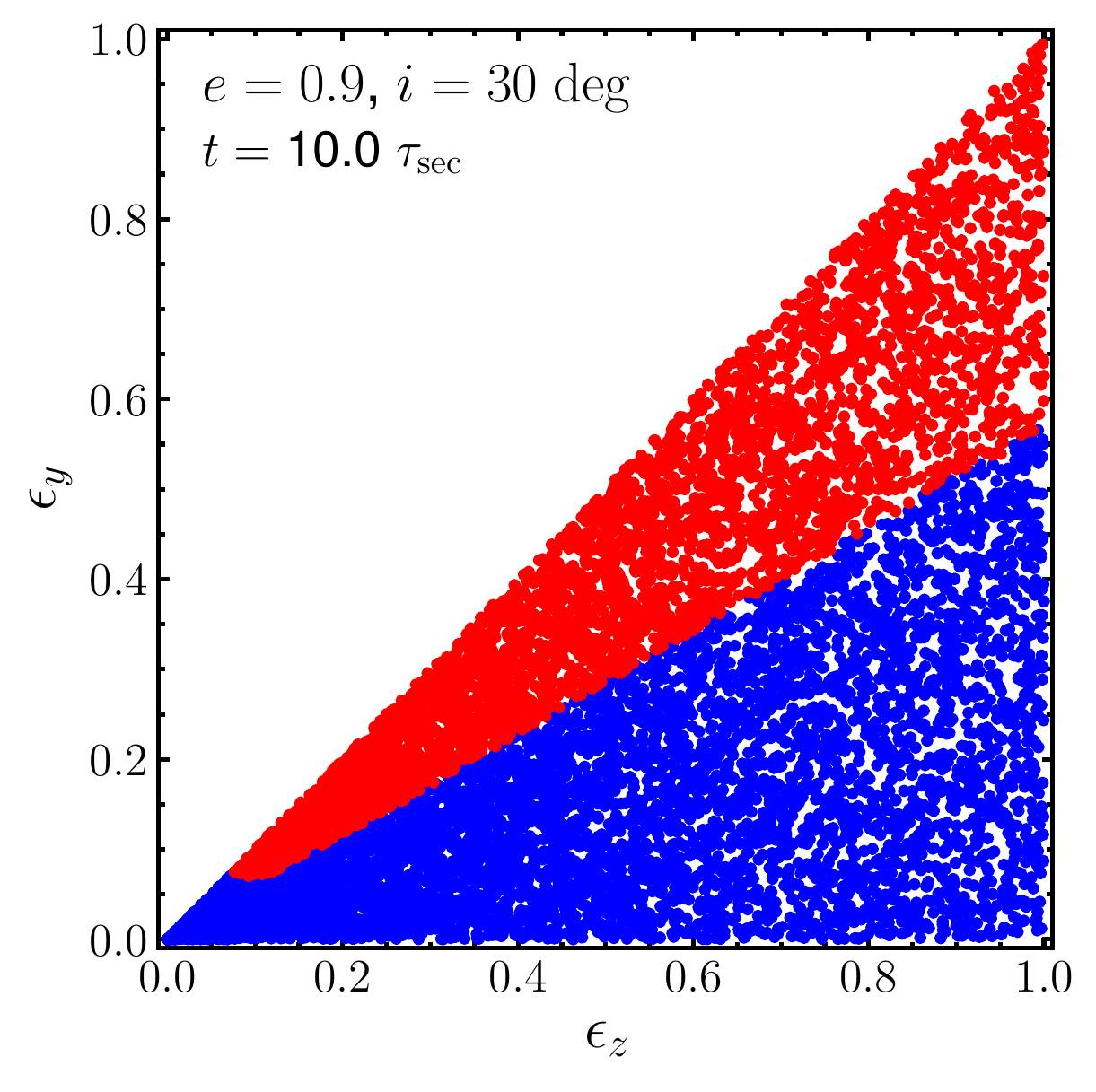}     
    \caption{Flip fraction for a single eccentric low inclination binary ($e=0.9$, $i=30\deg$) integrated in various triaxial potentials ($0<\epsy \lesssim \epsz$). We considered 5000 independent integrations and show the systems that flip in a time less than $t=1\tau_{\rm{sec}}$ (upper panel) and $t=10\tau_{\rm{sec}}$ (bottom panel). The binaries that flip are represented by the red filled circles. We observe that for sufficient large values of $\epsy\gtrsim 0.7$ the eccentric binary can flip in less than $1\tau_{\rm{sec}}$, while we need  
    $\epsy/\epsz\gtrsim 1/2$ for a flip to occur in less than  $10\tau_{\rm{sec}}$.}
    \label{fig:ecc_flip_vary_eps}
\end{figure}

In order to quantify the conditions for which the binary  angular momentum vector ${\bf j}$ flips with respect to the $\hat{\bf n}_z$ direction, we run a population model. 
We consider three different potentials which represent   triaxial merger remnants with varying levels of flattening. We classify the flattening of the potential into weak, intermediate and strong depending on the values of $\epsy$ and $\epsz$.
Thus, we consider three triaxial systems with increasing level of flattening,
where $\epsy$ and $\epsz$ where set to  $[0.05,0.2]$,   $[0.2,0.8]$  and $[0.6,0.8]$. Using equation~(\ref{arat}), these parameters correspond, respectively, to axis ratios $b/a$ and $c/a$ of $[{0.92},{0.68}]$, $[{0.8},{0.2}]$ and $[{0.5},{0.3}]$.

We uniformly sample the eccentricity $e \sim U(0, 1)$ and  the inclination $\cos(i) \sim U(-1, 1)$. The longitude of the ascending node and argument of pericentre are then sampled uniformly such that $\Omega, \omega \sim U(0, 2\pi)$. 
For each population synthesis run we consider a population of $10,000$ binaries and evolve all binaries until a time $t=50\tau_{\rm{sec}}$ where $\tau_{\rm{sec}}$ is defined by equation~\eqref{eq:secular_time}.
We classify a ``flip" when $j_z$ changes sign at any time during the integration, and a ``no flip" otherwise. We note that this is equivalent to the definition of the binary orbital flip presented in Section \ref{sec:results} where the binary orbit was found to flip with respect to the shortest axis of the galaxy (i.e., $\hat{\bf n}_z$). 
The results of the population synthesis model is presented in Fig.\ref{fig:pop_synth_flip}, which shows the initial eccentricity and inclination of the binaries. The red and blue filled circles show the initial parameters causing the binary to flip and not to flip, respectively. 

For the case of intermediate and weak levels of flattening (middle and right  panels in Fig.\ref{fig:pop_synth_flip}), we see that  $50.46\%$ and $53.57\%$ of binaries flip their orbital plane, respectively. We find that most binaries that are sufficiently inclined  ($70\deg \leq i \leq 110\deg$) reverse their sense of rotation, regardless of their initial eccentricity. 
Likewise, we  observe that most binaries that are sufficiently eccentric ($e>0.8$)  flip, regardless of their initial inclination.
This parameter space that is prone to flipping the binary orbital plane  is somewhat expected, as inclined and eccentric binaries require a small amount of torque in order to flip the $z$ component of the angular momentum.  These characteristics are also observed in the binaries integrated in the flatter potential (see the left panel in Fig.\ref{fig:pop_synth_flip}). The most noticeable difference is that the phase space region leading to binary flips is significantly larger, where the majority of binaries flip their orbital plane ($80\%$). We still observe  that the region of inclination phase space leading to flipping binaries increases with increasing eccentricity, where almost all binaries flip for $e>0.6$, irrespective of their initial inclination. These results suggest that any sufficiently eccentric and/or inclined binary  embedded in a triaxial potential is expected to flip if evolved for a sufficiently long time. We also considered integrations in an axisymmetric potential with $\epsilon_y=0$ and $\epsilon_z=0.2$. Unsurprisingly, in this latter case we observe no flip of the orbital plane irrespective of the initial conditions.

\subsection{Flip timescale}
\label{sec:flip_time}
In the bottom panels of Fig.\ref{fig:pop_synth_flip}, we plot the flip time distribution of the binaries integrated in the triaxial potentials introduced in the previous section. We find that for larger values of $\epsy$ and $\epsz$ the flip time distribution decreases and for the largest values (left panel) the median flip time is $\tau_{\rm{flip}}=2.0\tau_{\rm{sec}}$. This flip time is extremely quick, with a significant fraction of highly eccentric binaries flipping in $\tau_{\rm{flip}}<1.0 \tau_{\rm{sec}}$. To clearly demonstrate this rapid flip of the binary we considered an eccentric binary $e=0.9$ with $i=30 \deg$ which was integrated in a variety of triaxial potential strengths where we record the fraction of binaries that flip in less than $1$ and $10\tau_{\rm{sec}}$. We performed 5000 independent integrations with the results are illustrated in Fig.\ref{fig:ecc_flip_vary_eps} with the red and blue filled circles denoting the flipped and non-flipped binaries, respectively. Interestingly, we observe that for sufficiently large $\epsy \gtrsim 0.7$ all  binaries flip their orbital plane within a time $1\tau_{\rm{sec}}$, and  for 
any $\epsy/\epsz \gtrsim 2$ the binary can flip in less than $10\tau_{\rm{sec}}$.

In conclusion, our semi-analytical models predict that a sufficiently eccentric binary orbiting within a triaxial cusp can reverse its sense of
rotation over a  timescale  $\tau_{\rm flip} \sim \tau_{\rm sec}$.
Using the scaling introduced in Section~\ref{equa}, this approximately translates to a time $\sim 1.0$ in $N$-body units. The models presented here provide therefore a simple explanation of why in the $N$-body simulations any initially eccentric binary is observed to flip very rapidly at the time of binary formation.
Next, we derive a simple condition for the minimum eccentricity required for a flip to occur.

\subsection{Flip condition: analytical result for coplanar encounters}
\label{sec:flip_condition}

So far we have shown that a triaxial potential can cause the orbital plane of the binary to flip on a relatively short timescale. We also have shown that the flip fraction depends on the inclination and eccentricity of the binary prior to flipping, with a larger fraction of flips occurring for highly eccentric and inclined binaries. However, to better compare with our $N$-body simulations we would like to derive a flip condition in the case of a co-planar binary. To do this we follow an approach similar to \citet{bub2020}, where we derive a co-planar and Hamiltonian version $\mathcal{H}_{\rm{cop}}$ of our potential in Eq. \eqref{eq:tot_potential}.
To do this we expand this potential for the limit of zero inclination ($i \to 0$), thus setting  $j_{x}=j_{y}=e_{z}=0$. We further define the eccentricity vector on the orbital plane as $e_{x}=e\cos{\varpi}$ and $e_{y}=e\sin{\varpi}$, where $\varpi=\omega + \Omega$. Thus, in this coplanar limit, we write
\begin{align}
&\mathcal{H}_{\rm{cop}} = \pi G\tilde{\rho} \tilde{r}a
\left\{3-j_{}^2+ \epsy 
\left[j_{}^2+3e^2\sin^{2}{\varpi} \right]\right\}.
\label{eq:coplanar_hamiltonian}
\end{align}
To derive the flip condition we now equate the co-planar Hamiltonian at an initial ($e_0$, $\varpi_0$) and final condition, where the final condition takes the form $e\to 1$ (hence $j \to 0$),  $\varpi \to 0$.
From this we obtain the flip condition by equating the Hamiltonian states, which can be written as
\begin{equation}
    e_{0} \ge \sqrt{\frac{1-\epsy}{1-\epsy + 3\epsy\sin^2(\varpi_{0})}}.
    \label{eq:flip_condition}
\end{equation}
We observe that the flipping condition does not depend on $\epsz$. This is unsurprising given the coplanarity assumption where the torques from the $z-$component do not operate. We note that a similar condition in the context of secular three-body dynamics  was found by \citet{gongjie}, relating a minimum eccentricity to the level of non-axisymmetry of the potential from the outer orbit parametrized by $\epsilon_{\rm oct}=(a_{\rm in}/a_{\rm out}) e_{\rm out}/(1-e_{\rm out}^2)$ (e.g., \citealt{naoz_review}). Equivalently, $\epsilon_y$ measures the level of non-axisymmetry of the remnant's potential.

\begin{figure}
    \centering
    \includegraphics[width=0.9\columnwidth]{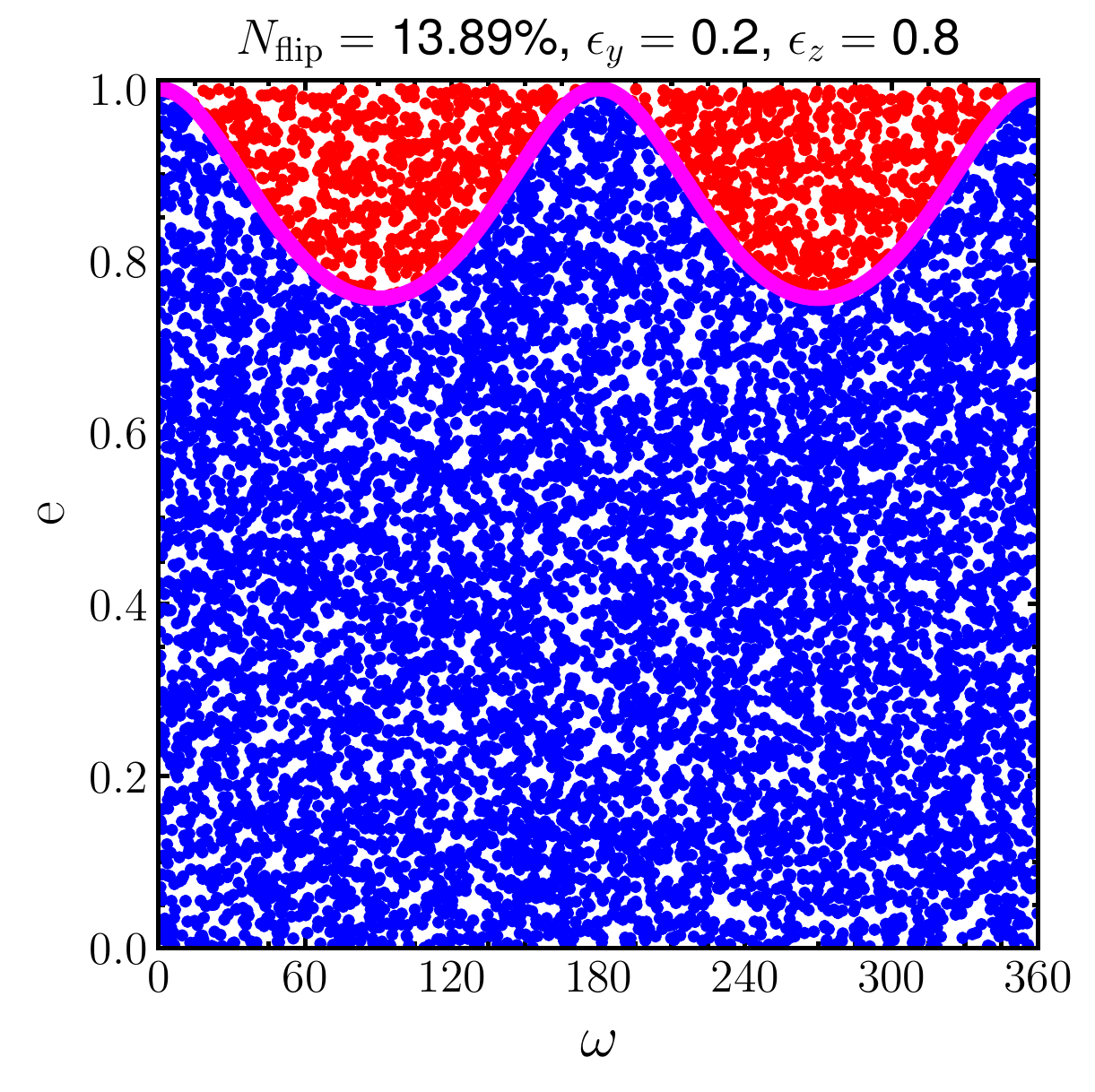} 
    \caption{The flip fraction in $e-\omega$ phase space of the 10000 binaries run in the population synthesis model. Binaries were run for $50$ secular times $\tau_{\rm{sec}}$, the blue filled circles represent the binaries that did not flip with respect to the $\hat{\bf n}_z$ and the red filled circles represent the binaries that flip. The solid magenta line is the flip condition derived from equation~\eqref{eq:flip_condition}. 
    The values for $\epsy$ and $\epsz$ are displayed at the top of the figure with the flip fraction $N_{\rm{frac}}$.
    The red and blue filled circles represent binaries that  flip and binaries that do not flip, respectively.}
    \label{fig:flip_condition}
\end{figure}

In order to test our flip condition, we ran a population synthesis using our averaged equations with $\epsy=0.2$ and $\epsz=0.8$ fixing $i=\Omega=0$, but varying $\omega$.
In Fig.\ref{fig:flip_condition} we plot the flip fraction in $e-\omega$ phase space (as $\Omega=0$, $\omega=\varpi$) and show  that our analytical co-planar flip condition (solid magenta line) is in excellent agreement, perfectly separating flip from non-flip systems.  
This simple model again explains why only very eccentric binaries are able to reverse their sense of rotation in the $N$-body models.

\begin{figure}
    \centering
    \includegraphics[width=1.0\columnwidth]{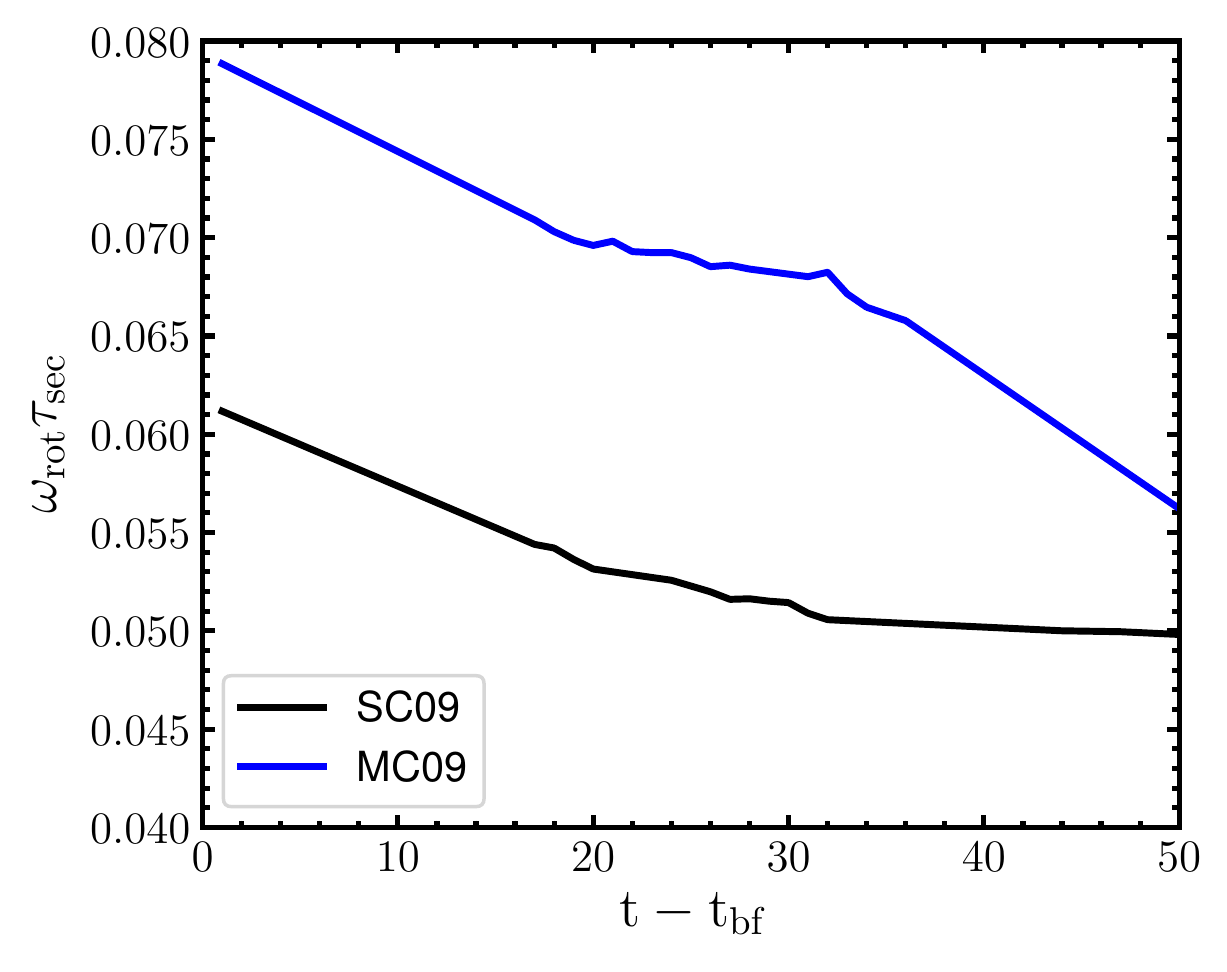} 
    \caption{The rotation $\omega_{\rm{rot}}\tau_{\rm{sec}}$ measured at the half mass radius for our $N$-body models where we observe the binary flip. We observe the decrease in rotation measure $\omega_{\rm{rot}}\tau_{\rm{sec}}$ after the time of binary formation for both $N$-body models.}
    \label{fig:omega_tau_half_mass}
\end{figure}

\subsection{Effect of figure rotation}
\label{sec:pop_syn_rot}

So far we have considered the orbital plane flip of the binary due to the triaxiality of the merger remnant. However, in the merging of two galaxies the merger remnant is expected to have some degree of rotation.
Previous studies have implied that non-zero streaming motion plays a part in the flipping of the orbital plane of a binary \citep{holleykhan2015,mirza2017,khan2019}, showing  preferential flipping for binaries embedded in  systems with an excess of stars that  
counter-rotate with the binary. It should be noted that the interpretation of this effect in previous studies has been different, where we consider the rotation of the global potential and not of the relative velocity between the binary SMBH and surrounding matter.
to a larger level of eccentricity enabling it to flip. 
However, the effect of figure rotation, where the entire figure of the galaxy rotates around some axis,  has never been investigated in this context.

We find that the figure of the $N$-body galaxy merger remnants rotates around the shortest axis of the galaxy (i.e., $\hat{\bf n}_z$) at a angular frequency $\omega_{\rm{rot}}$ that decreases with time. We compute $\omega_{\rm{rot}}$ by calculating the angular change in the direction of the longest-axis of the best fitting ellipsoid  (computed as in Section~\ref{bopf}) per time interval.
In Fig.\ref{fig:omega_tau_half_mass}  we show  $\omega_{\rm{rot}}\tau_{\rm{sec}}$ evaluated at the half-mass radius of the galaxy  as a function of time and after binary formation. 
 We see that the rotation at the time of binary formation is $\omega_{\rm{rot}}\tau_{\rm{sec}}\simeq 0.05$  and decreases slightly as a function of time for both models where we note the orbital plane flip (SC09 and MC09).

To consider the effect of rotation in our analytical treatment, 
 we add the rotational potential to that of the time-averaged potential where
 \citep[e.g.,][]{tremaine2014}:
\begin{equation}
\Phi_{\rm rot}=-\omega_{\rm{rot}}(GM_{\rm{bin}} a)^{1/2}\,{\bf j \cdot \hat{\bf n}}_z \ .
\end{equation}
This modifies the equations of motion such that

\begin{equation}\label{eq:dj_dt_rot}
\begin{split}
\frac{d {\bf {j}}}{d\tau}\bigg|_{\rm {rot}} &= \frac{d {\bf {j}}}{d\tau}\bigg|_{\rm {non,rot}} +  \left({\bf j} \times \hat{\bf n}_z\right)\omega_{\rm{rot}} \tau_{\rm sec}\\
\end{split}
\end{equation}

\begin{equation}\label{eq:de_dt_rot}
\begin{split}
\frac{d {\bf {e}}}{d\tau}\bigg|_{\rm {rot}} &= \frac{d {\bf {e}}}{d\tau}\bigg|_{\rm {non,rot}} +  \left({\bf e} \times \hat{\bf n}_z\right)\omega_{\rm{rot}}\tau_{\rm sec}\\
\end{split}
\end{equation}
where $\frac{d {\bf {j}}}{d\tau}|_{\rm {non,rot}}$ and $\frac{d {\bf {e}}}{d\tau}|_{\rm {non,rot}}$ are given by equations 
(\ref{eq:dj_dt}) and (\ref{eq:de_dt}), respectively.

\begin{figure*}
    \centering
    \includegraphics[width=2.25\columnwidth]{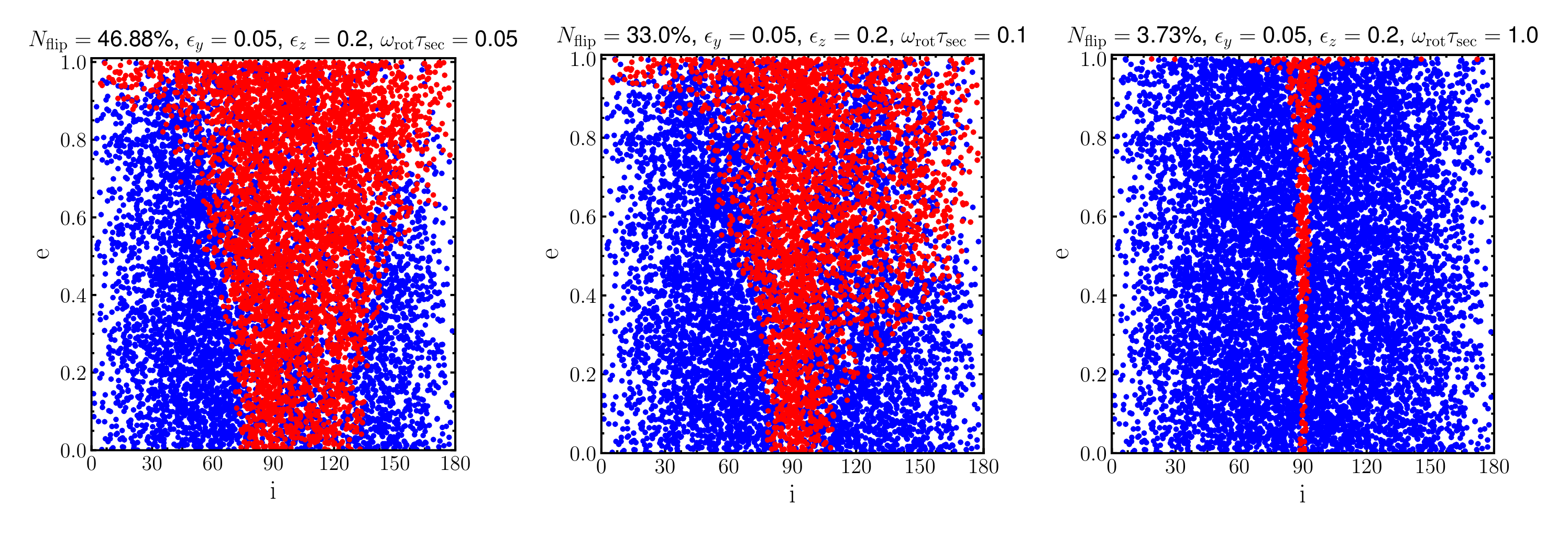} 
    \caption{The initial eccentricity and inclination phase space ($e-i$) showing the flip fraction $N_{\rm{frac}}$ where $\epsy=0.05$ and $\epsz=0.2$ for all of the population synthesis models each integrating 10000 binaries. We consider 3 different degrees of rotation of the remnant $\omega_{\rm{rot}} \tau_{\rm{sec}}=0.05$ (similar to that found in the $N$-body simulations, left panel), $\omega_{\rm{rot}} \tau_{\rm{sec}}=0.1$ (middle panel) and $\omega_{\rm{rot}} \tau_{\rm{sec}}=1.0$ (right panel). The red and blue filled circles represent the flipped and non-flipped binaries respectively.
    We observe that the binary orbital plane flip happens preferentially for counter-rotating binaries compared to co-rotating binaries.
    }
    \label{fig:pop_syn_rot}
\end{figure*}

We perform population synthesis models integrating binaries in the rotating time-averaged potential. Here we assume that the rotation is sufficiently slow that the orbit-averaged approximation remains valid. We consider three degrees of rotation
$\omega_{\rm{rot}} \tau_{\rm{sec}}=0.05, 0.1, 1.0$, with stronger levels of rotation for increasing values of $\omega_{\rm{rot}} \tau_{\rm{sec}}$. 
We set $\epsy=0.05$ and $\epsz=0.2$, the rest of the population synthesis parameters remain the same as described in Section \ref{sec:flip_frac}. We perform population synthesis models for positive values of $\omega_{\rm{rot}} \tau_{\rm{sec}}$, so that binaries with initially $i>90{\deg}$
are counter-rotating and those with $i<90{\deg}$ are co-rotating with the galaxy. The results for the population synthesis models are given in Fig.\ref{fig:pop_syn_rot} which shows the flip fraction in $e-i$ phase space.

Fig.\ref{fig:pop_syn_rot}, shows that binaries embedded in a counter-rotating system tend to flip more than binaries in a co-rotating potential. 
This characteristic can be inferred by observing that counter-rotating binaries ($i>90\deg$) have a larger phase space where binaries flip their orbital plane. 
This behaviour is most evident in the slow and intermediate levels of rotation (Fig.\ref{fig:pop_syn_rot}, upper and intermediate panels) where we observe a retrograde flip fraction $N_{\rm{flip,ret}}$ of $64.8\%$ and $63.2\%$, respectively. While this characteristic is less evident for the fast rotating case ($\omega_{\rm{rot}}\tau_{\rm{sec}}=1.0$), we still observe a preferential flip for retrograde binaries with $N_{\rm{flip,ret}}=52.3\%$.
We verified that these characteristics persist for different levels of rotation by running further populations synthesis models with $\omega_{\rm{rot}}\tau_{\rm{sec}}=0.01, 0.25$.
These results indicate that even for slow rotators such as those formed in the $N$-body simulations ($\omega_{\rm{rot}}\tau_{\rm{sec}}\simeq 0.05$), figure rotation should have a significant effect on  whether the orbital plane flip of the binary takes place.

From Fig.\ref{fig:pop_syn_rot} we also observe that increasing the strength of rotation (larger $\omega_{\rm{rot}}\tau_{\rm{sec}}$) leads to a smaller flip fraction where, for $\omega_{\rm{rot}} \tau_{\rm{sec}}=1.0$  rotation suppresses the orbital flip. This result is expected because a rapidly rotating triaxial potential would be effectively seen as an axisymmetric potential by the binary.

The analysis presented in this section, interestingly shows  that both the strength of triaxiality and the degree of figure rotation  affect whether the orbital plane flip of the binary takes place.
It demonstrates that  the phase space available for orbital plane flips in counter-rotating potentials is significantly larger than in co-rotating models, allowing less eccentric counter-rotating binaries to flip their orbital plane. 
 We note that while figure rotation should also affect the binary evolution observed in the $N-$body simulations,  it is difficult to discern its effect given the  small number of simulations at hand and that the binaries in these models are all initially co-rotating. It could  however be important and push the minimum eccentricity required for a flip to higher values than predicted by equation~(\ref{eq:flip_condition}).

\section{Conclusion}
\label{sec:conclusion}
In this study we have
performed $N$-body  simulations of the merger of equal-mass galaxies hosting a central SMBH. 
The binary that forms during the merging process is initially co-rotating with the surrounding galaxy. However, we show that sufficiently eccentric  binaries suddenly reverse their sense of rotation at the moment of the bound binary formation, leading to  a binary that is almost perfectly retrograde with respect to its galactic host.
This orbital plane flip is found to occur independently of the density profile slope  of the progenitor galaxies.

The counter-rotating binary that forms after the merger is expected to increase its eccentricity and evolve towards co-rotation as it exchanges energy and angular momentum with surrounding stars \citep[e.g.,][]{gualandris2012}. 
From our $N$-body simulations we confirm this prediction, and observe that after the binaries become counter-rotating they slowly re-align their sense of rotation with that of the host galaxy, enabling them to achieve extremely large eccentricities;
some reach $e\approx 0.99$ by the end of the numerical integration. 
Such eccentric binaries form in eccentric galaxy mergers, which naturally result in the formation of largely triaxial merger remnants \citep[see also][]{bortolas2018}. Using several realisations of the same model, we find that the orbital plane flip behaviour is not due to a specific choice of initial conditions or numerical resolution -- all binaries on very eccentric galactic orbits  flip almost exactly at the same time.

We suggest that the orbital flip of the binary is a result of the torque from the triaxial merger remnant. 
To investigate this, we consider an approximate analytical model for the evolution of a massive binary in a triaxial potential, and perform population synthesis models to explore the parameter space of initial conditions for binaries resulting in an orbital plane flip. The results of our analytical model support our hypothesis, showing that for sufficiently flattened  potentials an eccentric binary  can flip over within a few tens of its orbital period. In the case of co-planar systems, we derived an explicit orbital plane flip condition which demonstrates that flips are only experienced by co-planar SMBH binaries that have an initial eccentricity $e\gtrsim 0.75$, in agreement with the $N$-body models.

Finally, we find that {\it figure} rotation is present in the $N$-body merger models 
and include the effect of rotation in our analytical treatment. 
These new models show that binaries preferentially flip in counter-rotating systems compared to co-rotating systems.

From our numerical simulations we have shown that the process of forming counter-rotating binaries due to the orbital plane flip is a key ingredient in enabling the binaries to achieve very large eccentricities. 
These large eccentricities have a significant implication on the merger time-scale of SMBH binaries \citep[e.g.][]{Nasim2020}, leading to a faster coalescence as well as the possibility of detecting a residual eccentricity in the gravitational wave signal of such systems \citep{PS2010,ravi2014}. 

This rapid realignment of the orbital plane of the SMBH binary may also have significant observational implications. It has been shown that the direction of the spin axis of the resultant SMBH from the binary merger is affected by the orbital plane orientation of the binary prior to coalescence \citep{merritt2002}, which, in turn, determines the orientation
of the accretion disk around the resultant SMBH via the
Bardeen-Peterson effect \citep{bardeen_petterson1975}
and similarly the direction of the radio jet from AGNs.

We conclude by briefly comparing to previous work. \citet{mirza2017} and \citet{khan2019} considered rotating, axisymmetric galaxy models and
noted a similar SMBH binary orbital plane flip, but only for initially counter-rotating systems, leading to co-rotating binaries. 
They attributed this to the preferential ejection of prograde stars by the binary \citep{gualandris2012}. 
As noted above, this cannot be the reason for the flip in our models because our binaries are initially co-rotating with the galaxy.
This characteristic combined with the fact that we observe the rapid flip of the binary orbital plane implies that the flip behaviour we observe cannot be due to the mechanism proposed by \citet{dotti2006}.
Although a direct comparison with the models
 in \citet{mirza2017} and \citet{khan2019} is required to draw firmer conclusions,
we propose here that  the flip observed by them  might also be caused by the torque from the non-axisymmetric mass distribution surrounding the binary. The idea that the two mechanisms might be the same is supported by three  similarities: (i) their SMBH binaries also flip suddenly at the time of binary formation; (ii) only their most eccentric binaries flip; and (iii) as noted in \citet{holleykhan2015}, their galaxy models exhibit a more triaxial shape inside the radius of influence of the SMBHs.
 Thus,  it may well be that
this inner region in their models was sufficiently triaxial to allow the orbital plane flip to occur. 
In this picture, 
the fact that \citet{mirza2017} and \citet{khan2019}  did not observe a flip
in their initially co-rotating binaries could be explained by their smaller initial eccentricity, rather than by their sense of rotation. 
However it is worth noting that the larger scale studies \citep{dotti2006,dotti2007} show angular momentum flips resulting in prograde systems, albeit via a fundamentally different mechanism, which are associated with the local streaming velocities in systems with negligible triaxiality.

\section*{Acknowledgements}
 CP acknowledges support from the Bart J. Bok fellowship at Steward Observatory.
  FD acknowledges support from PCTS and Lyman Spitzer Jr fellowships.
FA acknowledges support from a Rutherford fellowship
(ST/P00492X/1) from the Science and Technology Facilities Council. 

\section*{Data availability}
The data underlying this article will be shared on a reasonable request to the corresponding author.


\bibliographystyle{mnras}
\bibliography{main}

\appendix
\section{equations of motion}\label{ap:equations_of_motion}
To derive the orbit-averaged equations of motion we follow a similar approach to \citep{PA2017} where the potential per unit mass defined in equation~(\ref{eq:triaxial_potential}) is
\begin{equation}
\begin{split}
{\Phi}({\bf r})&=\frac{4\pi G}{(3-\gamma)(2-\gamma)}
\tilde{\rho} \tilde{r}^2\left(\frac{r}{\tilde{r}}\right)^{2-\gamma}\\
& \times\left[1+
\epsilon_z\frac{(\hat{\bf n}_z\cdot {\bf r})^2}{r^2}+
\epsilon_y\frac{(\hat{\bf n}_y\cdot {\bf r})^2}{r^2}\right].
\end{split}
\end{equation}
Firstly, we perform the averaging the terms that appear on the triaxial potential.
Assuming that ${\bf r}$ follows a Keplerian orbit,  we can average an arbitrary function $f(\mathbf{r})$ over the orbit of the binary as
\begin{eqnarray}
\label{eq:average_of_function}
    \langle f(\mathbf{r}) \rangle = \frac{(1 - e^2)^{3/2}}{2 \pi} \int_0^{2 \pi} \frac{d \phi}{(1 + e \cos \phi)^2} f(r, \phi).
\end{eqnarray}

For the case of non-spherical potential, the following results are useful for averaging terms.
\begin{subequations}
\begin{align}
r_{}^{-\gamma}(\hat{\bf n}_i\cdot {\bf r}_{})^2= 
r_{}^{2-\gamma}(\hat{\bf n}_i\cdot {\hat{\bf r}}_{})^2 = 
r_{}^{2-\gamma}\cos^2 \phi 
\left(\hat{\bf n}_i\cdot {\bf \hat{e}_{}}\right)^2 \notag \\
+2r_{}^{2-\gamma}\sin \phi\cos \phi \left(\hat{\bf n}_i\cdot {\bf \hat{e}}_{}\right)
\left(\hat{\bf n}_i\cdot {\bf \hat{q}}_{}\right) 
+
r_{}^{2-\gamma}\sin^2 \phi \left(\hat{\bf n}_i\cdot {\bf \hat{q}}_{}\right)^2,
\label{eq:useful_relation}
\end{align}
\begin{align}
    \left(\hat{\bf n}_i\cdot {\bf \hat{q}}_{}\right) \left(\hat{\bf n}_j\cdot {\bf \hat{q}}_{}\right) = \delta_{ij} - \left(\hat{\bf n}_i\cdot {\bf \hat{e}}_{}\right) \left(\hat{\bf n}_j\cdot {\bf \hat{e}}_{}\right) - \left(\hat{\bf n}_i\cdot {\bf \hat{j}}_{}\right) \left(\hat{\bf n}_j\cdot {\bf \hat{j}}_{}\right) 
\end{align}

\end{subequations}

Using the fact that $r = a(1-e^2)/(1+e\cos{\phi})$ we average over the azimuthal angle $\phi$ which gives us the total averaged potential which takes the form 
\begin{align}
&\langle\Phi\rangle =\pi G\tilde{\rho} \tilde{r}a 
\left\{3-j_{}^2+\sum_{i=y,z}  \epsilon_i 
\left[j_{}^2+3\left(\hat{\bf n}_i\cdot {\bf {e}}_{}\right)^2
-\left(\hat{\bf n}_i\cdot {\bf {j}}_{}\right)^2 \right]\right\}.
\end{align}

From the time-averaged potential we can derive the Milankovitch equations of motion using the relation \citep[e.g.][]{tremaine2014}
\begin{align}
    \frac{d {\bf {j}}}{dt} &= -\frac{1}{\sqrt{GM_\mathrm{bin}a}} ({\bf {j}} \times \nabla_{{\bf {j}}} \langle \Phi \rangle + {\bf {e}} \times \nabla_{{\bf {e}}} \langle \Phi \rangle ) \\
    \frac{d {\bf {e}}}{dt} &= -\frac{1}{\sqrt{GM_\mathrm{bin}a}} ({\bf {j}} \times \nabla_{{\bf {e}}} \langle \Phi \rangle + {\bf {e}} \times \nabla_{{\bf {j}}} \langle \Phi \rangle ) .
    \label{eq:milankovitch}
\end{align}
Applying the relations given in equation \eqref{eq:milankovitch}, we obtain our equations of motion
\begin{equation}
\begin{split}\label{eq:ap_dj_dt}
\frac{d {\bf {j}}}{dt} &= -\frac{\pi \sqrt{G}\tilde{\rho} \tilde{r} \sqrt{a}}{\sqrt{M_\mathrm{bin}}}  \\
&\Bigg\{\sum_{i=y,z}  \epsilon_i 
\left[-2\left(\hat{\bf n}_i\cdot {\bf {j}}_{}\right) \left({\bf j} \times \hat{\bf n}_i\right) +6\left(\hat{\bf n}_i\cdot {\bf {e}}_{}\right) \left({\bf e} \times \hat{\bf n}_i\right) \right]\Biggl\}. \\
\end{split}
\end{equation}

\begin{equation}
\begin{split}\label{eq:ap_de_dt}
\frac{d {\bf {e}}}{dt} &= -\frac{\pi \sqrt{G}\tilde{\rho} \tilde{r} \sqrt{a}}{\sqrt{M_\mathrm{bin}}}  \\
&\Bigg\{\sum_{i=y,z}  \epsilon_i 
\left[-2\left(\hat{\bf n}_i\cdot {\bf {j}}_{}\right) \left({\bf e} \times \hat{\bf n}_i\right) +6\left(\hat{\bf n}_i\cdot {\bf {e}}_{}\right) \left({\bf j} \times \hat{\bf n}_i\right) \right]\Biggl\}. \\
\end{split}
\end{equation}

\bsp	
\label{lastpage}
\end{document}